\documentclass[12pt]{iopart}
\usepackage{iopams}  
\usepackage{graphicx}
\usepackage{amsfonts}
\usepackage{latexsym}
\usepackage{amssymb} 
\usepackage{verbatim}
\usepackage{amsthm}

\newcommand{\NN}{{{\mathbb N}}}

\newcommand{\RF}{{{\mathbb R}}}

\newtheorem{theorem}{Theorem}[section]

\newtheorem{proposition}{Proposition}[section]

\newtheorem{conjecture}{Conjecture}[section]
\begin{document}
\title[
Recursion structure in the definition of
gauge-invariant variables ...
]
{
  Recursive structure in the definitions of
  gauge-invariant variables for any order perturbations
}
\author{
  Kouji Nakamura
}
\address{
  TAMA Project, Optical and Infrared Astronomy Division,\\
  National Astronomical Observatory of Japan,\\
  2-21-1, Osawa, Mitaka, Tokyo 181-8588, Japan
}
\ead{kouji.nakamura@nao.ac.jp}
\begin{abstract}
  The construction of gauge-invariant variables for any order 
  perturbations is discussed.
  Explicit constructions of the gauge-invariant variables for
  perturbations to 4th order are shown.
  From these explicit constructions, the recursive structure
  in the definitions of gauge-invariant variables for
  any order perturbations is found.
  Through this recursive structure, the correspondence with the
  fully non-linear exact perturbations is briefly discussed.
\end{abstract}


\section{Introduction}
\label{sec:K.Nakamura-2014-1}


Higher-order perturbation theory is one of topical subjects in
the recent research on general relativity and have very wide
applications: cosmological
perturbations~\cite{Tomita-1967-Non-Gaussianity}; black hole 
perturbations~\cite{Gleiser-Nicasio-2000}; and perturbations of
stars~\cite{Kojima-1997}. 
However, the ``gauge issues'' in higher-order perturbations are
very delicate in spite of their wide applications. 
Therefore, it is worthwhile to discuss the higher-order
perturbation theory in general relativity from general point of
view.
Due to this motivation, we have been formulating the higher-order
perturbation theory in general relativity through a
gauge-invariant
manner~\cite{kouchan-gauge-inv,kouchan-second,kouchan-decomp}
and applied our formulation to cosmological
perturbations~\cite{Nakamura:2010yg}.
These works are mainly concerning about the second-order
perturbations except for Ref.~\cite{kouchan-gauge-inv}.
In this paper, we discuss the ``gauge issues'' for any order
perturbations.


General relativity is a theory based on general covariance and
the notion of ``gauge'' is introduced in the theory due to this 
general covariance.
In particular, in general-relativistic perturbations, {\it the
  second kind gauge} appears in perturbations as Sachs pointed 
out~\cite{R.K.Sachs-1964}. 
In general-relativistic perturbation theory, we usually treat 
one-parameter family of spacetimes 
$\{({\cal M}_{\lambda},Q_{\lambda})|\lambda\in[0,1]\}$ to
discuss differences between the background spacetime 
$({\cal M}_{0},Q_{0})=({\cal M}_{\lambda=0},Q_{\lambda=0})$ and
the physical spacetime $({\cal M}_{\lambda=1},Q_{\lambda=1})$. 
Here, $\lambda$ is the infinitesimal parameter for
perturbations, ${\cal M}_{\lambda}$ is a spacetime manifold for
each $\lambda$, and $Q_{\lambda}$ is the collection of the
tensor fields on ${\cal M}_{\lambda}$.
Since each ${\cal M}_{\lambda}$ is different manifold, we have
to introduce the point-identification map ${\cal X}_{\lambda}$
$:$ ${\cal M}_{0}\mapsto{\cal M}_{\lambda}$ to compare the
tensor field on different manifolds.
This point-identification is {\it the gauge choice of the second
  kind}.
Since we have no guiding principle to choose the identification
map ${\cal X}_{\lambda}$ due to the general covariance, we may
choose a different point-identification ${\cal Y}_{\lambda}$
from ${\cal X}_{\lambda}$. 
This degree of freedom of the choice is {\it the gauge degree of
  freedom of the second kind}.
{\it The gauge-transformation of the second kind} is a change of
this identification map.
We note that this second-kind gauge is different notion of the
degree of freedom of coordinate choices on a single manifold,
which is called {\it the gauge of the first kind}.
Henceforth, we concentrate only on gauge of the second kind and
we call this second kind gauge as {\it gauge} for short.


Once we introduce the gauge choice ${\cal X}_{\lambda}$
$:$ ${\cal M}_{0}\mapsto{\cal M}_{\lambda}$, we can compare the
tensor fields on different manifolds $\{{\cal M}_{\lambda}\}$
and {\it perturbations} of a tensor field $Q_{\lambda}$ are
represented by the difference
\begin{eqnarray}
  \label{eq:perturbation-X-gauge-full}
  {\cal X}_{\lambda}^{*}Q_{\lambda} - Q_{0},
\end{eqnarray}
where ${\cal X}_{\lambda}^{*}$ is the pull-back induced by the
gauge choice ${\cal X}_{\lambda}$ and $Q_{0}$ is the background
value of the variable $Q_{\lambda}$.
We note that this representation of perturbations are completely
depends on gauge choice ${\cal X}_{\lambda}$.
If we change the gauge choice from ${\cal X}_{\lambda}$ to
${\cal Y}_{\lambda}$, the pulled-back variable of $Q_{\lambda}$
represented by the different representation
${\cal Y}_{\lambda}^{*}Q_{\lambda}$. 
These different representations are related to the
gauge-transformation rules as
\begin{eqnarray}
  \label{eq:gauge-trans-X-to-Y-full}
  {\cal Y}_{\lambda}^{*}Q_{\lambda}
  =
  \Phi^{*}_{\lambda} {\cal X}_{\lambda}^{*}Q_{\lambda},
\end{eqnarray}
where
\begin{eqnarray}
  \label{eq:Phi-def-Xinv-Y-full}
  \Phi_{\lambda} := ({\cal X}_{\lambda})^{-1}\circ{\cal Y}_{\lambda}
\end{eqnarray}
is a diffeomorphism on ${\cal M}_{0}$.


In the perturbative approach, we treat the perturbation 
${\cal X}_{\lambda}^{*}Q_{\lambda}$ through the Taylor series
with respect to the infinitesimal parameter $\lambda$ as
\begin{eqnarray}
  \label{eq:XQ-Tayler-expansion}
  {\cal X}_{\lambda}^{*}Q_{\lambda}
  =
  \sum_{n=0}^{k} \frac{\lambda^{n}}{k!} {}^{(n)}_{\;{\cal X}}\!Q
  +
  O(\lambda^{k+1})
  ,
\end{eqnarray}
where ${}^{(k)}_{\;{\cal X}}\!Q$ is the representation
associated with the gauge choice ${\cal X}_{\lambda}$ of the
$k$th order perturbation of the variable $Q_{\lambda}$ with its 
background value ${}^{(0)}_{\;{\cal X}}\!Q=Q_{0}$.
Similarly, we can have the representation of the perturbation of
the variable $Q_{\lambda}$ under the gauge choice 
${\cal Y}_{\lambda}$ which is different from ${\cal X}_{\lambda}$ 
as mentioned above.
Since these different representations are related to the
gauge-transformation rule (\ref{eq:gauge-trans-X-to-Y-full}),
the order-by-order gauge-transformation rule between $n$th-order
perturbations ${}^{(n)}_{\;{\cal X}}\!Q$ and ${}^{(n)}_{\;{\cal Y}}\!Q$
are given from the Taylor expansion of the gauge-transformation
rule (\ref{eq:gauge-trans-X-to-Y-full}).


Since $\Phi_{\lambda}$ is constructed by the product of
diffeomorphisms, $\Phi_{\lambda}$ is not given by an
exponential 
map~\cite{kouchan-gauge-inv,Nakamura:2010yg,M.Bruni-S.Soonego-CQG1997,S.Sonego-M.Bruni-CMP1998}, 
in general.
For this reason, Sonego and
Bruni~\cite{S.Sonego-M.Bruni-CMP1998} introduced the notion of a
{\it knight diffeomorphism}.
The knight diffeomorphism, which are generated by many
generators, includes wider class of diffeomorphisms than
exponential maps which are generated by a single vector field. 
This knight diffeomorphism is suitable for our order-by-order
arguments on the gauge issues of general-relativistic
higher-order perturbations.
Sonego and Bruni also derived the gauge-transformation rules for
any order perturbations.


The purpose of this paper is to point out the recursive structure
in the definition of the gauge-invariant variables for the
$n$th-order perturbations. 
We use the gauge-transformation rules for perturbations derived
by Sonego and Bruni.
We demonstrate the explicit constructions of gauge-invariant
variables to 4th order.
From these explicit constructions, we found the recursive
structure in the definitions of the gauge-invariant variables for
the $n$th-order perturbations based on algebraic recursion
relations
(Conjecture~\ref{conjecture:nth-order-gauge-transformation-conjecture})
and the decomposition of the linear metric perturbation into its
gauge-invariant and gauge-variant parts
(Conjecture~\ref{conjecture:decomposition-conjecture}).


The organization of this paper is as follows.
In section~\ref{sec:K.Nakamura-2014-2}, we review the knight
diffeomorphism introduced by Sonego and
Bruni~\cite{S.Sonego-M.Bruni-CMP1998} and gauge-transformation
rules derived them.
In section~\ref{sec:K.Nakamura-2014-3}, we examine the construction
of gauge-invariant variables to 4th-order perturbations.
These constructions are based on the conjecture which state that
we already know how to construct gauge-invariant variables for
linear-order metric perturbation
(Conjecture~\ref{conjecture:decomposition-conjecture}).
In section~\ref{sec:K.Nakamura-2010-4}, we discuss the recursive 
structure in the definitions of gauge-invariant variables for
$n$th-order perturbations.
Although this discussion is based on the conjecture for an
algebraic identities
(Conjecture~\ref{conjecture:nth-order-gauge-transformation-conjecture}), 
this algebraic identities are confirmed to 4th order
perturbation within this paper in
section~\ref{sec:K.Nakamura-2014-3}.
In section~\ref{sec:K.Nakamura-2010-5}, we discuss the
application of our formulae to cosmological perturbations as an
example. 
The final section (section~\ref{sec:K.Nakamura-2010-6}) is
devoted to the summary and discussions.


\section{Gauge-transformation rules of higher-order perturbations}
\label{sec:K.Nakamura-2014-2}


In this section, we briefly review a representation of
diffeomorphism proposed by Sonego and
Bruni~\cite{S.Sonego-M.Bruni-CMP1998}, which called a knight
diffeomorphism and the gauge-transformation rules for
$n$th-order perturbations.
In gauge-invariant perturbation theories, we may concentrate on
the diffeomorphism on the background spacetime ${\cal M}_{0}$.
However, in this section, we denote the spacetime manifold by
${\cal M}$ instead of ${\cal M}_{0}$, since our arguments are
not restricted to a specific background spacetime ${\cal M}_{0}$
in perturbation theories.


\subsection{Knight diffeomorphism}
\label{sec:K.Nakamura-2014-2.1}


Let $\phi^{(1)},...,\phi^{(k)}$ be exponential maps on 
${\cal M}$ which are generated by the vector fields
$\xi_{(1)},...,\xi_{(k)}$, respectively.
From these exponential maps, we can define a new one-parameter
family of diffeomorphisms $\Psi_{\lambda}^{(k)}$ on ${\cal M}$,
whose action is given by
\begin{eqnarray}
  \label{eq:S.Sonego-M.Bruni-1998-2.1}
  \Psi_{\lambda}^{(k)}
  :=
  \phi_{\lambda^{k}/k!}^{(k)}
  \circ\cdots\circ
  \phi_{\lambda^{2}/2}^{(2)}\circ\phi_{\lambda}^{(1)}
  .
\end{eqnarray}
$\Psi_{\lambda}^{(k)}$ displaces a point of ${\cal M}$, a
parameter interval $\lambda$ along the integral curve of
$\xi_{(1)}$, then an interval $\lambda^{2}/2$ along the integral
curve of $\xi_{(2)}$, and so on.
For this reason, Sonego and Bruni called $\Psi_{\lambda}^{(k)}$,
with a chess-inspired terminology, a {\it knight diffeomorphism
  of rank $k$}.
The vector fields $\xi_{(1)},...,\xi_{(k)}$ are called the
generators of $\Psi_{\lambda}^{(k)}$. 
The notion of this knight diffeomorphism is useful in
perturbation theories in the theories of gravity with general
covariance.
The reason of this usefulness is in the fact that any $C^{k}$
one-parameter family $\Phi_{\lambda}$ of diffeomorphisms can
always be approximated by a family of knights diffeomorphism of
rank $k$.
Actually, in \cite{S.Sonego-M.Bruni-CMP1998}, Sonego and Bruni
showed the following theorem:


\begin{theorem}
  \label{theorem:S.Sonego-M.Bruni-1998-1}
  Let ${\cal D}$ be an appropriate open set in
  $\{\lambda\}\times{\cal M}$ which includes
  $\{0\}\times{\cal M}$, $\lambda\in\RF$, and 
  $\Phi_{\lambda}:{\cal D}\rightarrow{\cal M}$ be a $C^{k}$
  one-parameter family of diffeomorphisms. 
  Then, there exists a set of exponential maps
  $\{\phi^{(1)},...,\phi^{(k)}\}$ on ${\cal M}$ such that, up to
  the order $\lambda^{k+1}$, the action of $\Phi_{\lambda}$ is
  equivalent to the one of the $C^{k}$ knight diffeomorphisms
  \begin{eqnarray}
    \label{eq:S.Sonego-M.Bruni-1998-2.2}
    \Phi_{\lambda}
    =
    \Psi_{\lambda}^{(k)}
    +
    O(\lambda^{k+1})
    =
    \phi^{(k)}_{\lambda^{k}/k!}
    \circ
    \cdots
    \circ
    \phi^{(2)}_{\lambda^{2}/2!}
    \circ
    \phi_{\lambda}^{(1)}
    +
    O(\lambda^{k+1})
    .
  \end{eqnarray}
\end{theorem}


If $\Phi$ and $\Psi$ are two diffeomorphisms of 
${\cal M}$ such that $\Phi^{*}f=\Psi^{*}f$ for every function
$f$, it follows that $\Phi\equiv\Psi$.
In order to show that a family of knight $\Psi_{\lambda}^{(k)}$
approximates any one-parameter family of diffeomorphisms
$\Phi_{\lambda}$ up to the $(k+1)$th order, it is sufficient to
prove that $\Psi_{\lambda}^{(k)*}f$ and $\Phi_{\lambda}^{*}f$
differ by a function that is $O(\lambda^{k+1})$ for all $f$.
We can always generalize the above approximation
property of the action of a knight diffeomorphism
$\Psi_{\lambda}^{(k)*}$ for an arbitrary function to that of the
action for an arbitrary tensor field.
For this reason, Sonego and Bruni concentrated on
Taylor-expansion of the pull-back
$\Psi_{\lambda}^{(k)*}f=\phi^{(1)*}_{\lambda}\phi_{\lambda^{2}/2}^{(2)*}\cdots\phi_{\lambda^{k}/k!}^{(k)*}f$
of a knight diffeomorphism for an arbitrary smooth function $f$
on ${\cal M}$.
Then they showed the following proposition:


\begin{proposition}
  \label{proposition:S.Sonego-M.Bruni-1998-4}
  Let $\Phi_{\lambda}$ be a one-parameter family of
  diffeomorphisms, and $T$ a tensor field such that
  $\Phi_{\lambda}^{*}T$ is of class $C^{k}$.
  Then, $\Phi_{\lambda}^{*}T$ can be expanded around $\lambda=0$
  as
  \begin{eqnarray}
    \label{eq:S.Sonego-M.Bruni-1998-3.8}
    \Phi_{\lambda}^{*}T
    =
    \sum_{l=0}^{k} \lambda^{l} \sum_{\{j_{i}\}\in J_{l}}
    {\cal C}_{l}(\{j_{i}\})
    {\pounds}_{\xi_{(1)}}^{j_{1}}\cdots{\pounds}_{\xi_{(l)}}^{j_{l}} T
    +
    O(\lambda^{k+1}).
  \end{eqnarray}
  Here, 
  $\displaystyle J_{n}:=\{\{j_{i}\}|\forall i\in\NN,j_{i}\in\NN,\;\mbox{s.t.}\;\sum_{i=1}^{\infty}ij_{i}=n\}$ 
  defines the set of indices over which one has to sum in order
  to obtain the $n$th-order term, 
  \begin{eqnarray}
    {\cal C}_{l}(\{j_{i}\})
    :=
    \prod_{i=1}^{l}
    \frac{1}{(i!)^{j_{i}}j_{i}!}
    ,
  \end{eqnarray}
  and $O(\lambda^{k+1})$ is a remainder with
  $O(\lambda^{k+1})/\lambda^{k}\rightarrow 0$ in the limit
  $\lambda\rightarrow 0$.
\end{proposition}


Here, we note that the expression of the right-hand side of
equation (\ref{eq:S.Sonego-M.Bruni-1998-3.8}) is just the form
of the Taylor-expansion of the right-hand side of equation
(\ref{eq:S.Sonego-M.Bruni-1998-2.1}).
From this fact, the
proposition~\ref{proposition:S.Sonego-M.Bruni-1998-4}, and the
fact that $\Phi\equiv\Psi$ if $\Phi$ and $\Psi$ are two
diffeomorphisms such that $\Phi^{*}f=\Psi^{*}f$ for every 
function $f$, we reach to the assertion of
Theorem~\ref{theorem:S.Sonego-M.Bruni-1998-1}. 
Therefore, we may regard that the Taylor-expansion
(\ref{eq:S.Sonego-M.Bruni-1998-3.8}) in 
Proposition~\ref{proposition:S.Sonego-M.Bruni-1998-4} is the
most general expression of the pull-back of diffeomorphism on
${\cal M}$ and it is sufficient at least when we concentrate on
perturbation theories. 
We also note that the properties of the set $J_{n}$ of integers
are discussed in \ref{sec:K.Nakamura-2014-appendix}.


\subsection{Gauge-transformation rule for the $n$th-order perturbations}
\label{sec:K.Nakamura-2014-2.2}


Through the notion of the knights diffeomorphism in the
previous section, we derive the gauge-transformation rules
for the $n$th-order perturbations.
As mentioned in section \ref{sec:K.Nakamura-2014-1}, the
gauge-transformation rule between the pulled-back variables
${\cal Y}_{\lambda}^{*}Q_{\lambda}$ and 
${\cal X}_{\lambda}^{*}Q_{\lambda}$ is given by
(\ref{eq:gauge-trans-X-to-Y-full}).
In perturbation theories, we always use the Taylor-expansion of
these variables as in equation (\ref{eq:XQ-Tayler-expansion}).
To derive the order-by-order gauge-transformation rule for the
$n$th-order perturbation, we have to know the form of the
Taylor-expansion of the pull-back $\Phi^{*}_{\lambda}$ of
diffeomorphism. 
Then, we use the general expression
(\ref{eq:S.Sonego-M.Bruni-1998-3.8}) of the Taylor expansion of
diffeomorphisms in
Proposition~\ref{proposition:S.Sonego-M.Bruni-1998-4} by Sonego
and Bruni.
Substituting equations (\ref{eq:S.Sonego-M.Bruni-1998-3.8}) and 
(\ref{eq:XQ-Tayler-expansion}) into equation
(\ref{eq:gauge-trans-X-to-Y-full}), we obtain the order-by-order
expression of the gauge-transformation rules between the
perturbative variables ${}^{(n)}_{\;{\cal X}}\!Q$ and 
${}^{(n)}_{\;{\cal Y}}\!Q$ as
\begin{eqnarray}
  {}^{(n)}_{\;{\cal Y}}\!Q
  -
  {}^{(n)}_{\;{\cal X}}\!Q
  &=&
  \sum_{l=1}^{n}
  \frac{n!}{(n-l)!}
  \sum_{\{j_{i}\}\in J_{l}}
  {\cal C}_{l}(\{j_{i}\})
  {\pounds}_{\xi_{(1)}}^{j_{1}}\cdots{\pounds}_{\xi_{(l)}}^{j_{l}}
  {}^{(n-l)}_{\;\;\;\;\;\;{\cal X}}\!Q
  .
  \label{eq:Sonego-Bruni-1998-5.1}
\end{eqnarray}
The order-by-order gauge-transformation rule
(\ref{eq:Sonego-Bruni-1998-5.1}) gives a complete description of
the gauge behavior of perturbations at any order.


\section{Definitions of gauge-invariant variables to 4th-order perturbations}
\label{sec:K.Nakamura-2014-3}


Inspecting the gauge-transformation rule
(\ref{eq:Sonego-Bruni-1998-5.1}), we define gauge-invariant
variables for metric perturbations and for perturbations of
arbitrary tensor fields.
Since the definitions of gauge-invariant variables for
perturbations of arbitrary tensor fields are trivial if we
accomplish the separation of the metric perturbations into their
gauge-invariant and gauge-variant parts.
Therefore, we may concentrate on the metric perturbations.


First, we consider the metric $\bar{g}_{ab}$ on the physical
spacetime $({\cal M}_{\lambda=1},Q_{\lambda=1})$.
We expand the pulled-back metric
${\cal X}_{\lambda}^{*}\bar{g}_{ab}$ to ${\cal M}_{0}$ through a
gauge choice ${\cal X}_{\lambda}$ as
\begin{eqnarray}
  {\cal X}^{*}_{\lambda}\bar{g}_{ab}
  &=&
  \sum_{n=0}^{k}\frac{\lambda^{n}}{n!} \; {}_{\;{\cal X}}^{(n)}\!g_{ab}
  + O(\lambda^{k+1})
  \label{eq:metric-expansion}
  .
\end{eqnarray}
where $g_{ab}:={}_{\;{\cal X}}^{(0)}\!g_{ab}$ is the metric on the
background spacetime ${\cal M}_{0}$. 
Of course, the expansion (\ref{eq:metric-expansion}) of the
metric depends entirely on the gauge choice 
${\cal X}_{\lambda}$.
Nevertheless, henceforth, we do not explicitly express the index
of the gauge choice ${\cal X}_{\lambda}$ if there is no
possibility of confusion.


In \cite{kouchan-gauge-inv}, we proposed a procedure to
construct gauge-invariant variables for higher-order
perturbations.
Our starting point to construct gauge-invariant variables was
the following conjecture for the linear-metric perturbation
$h_{ab}:={}^{(1)}\!g_{ab}$:


\begin{conjecture}
  \label{conjecture:decomposition-conjecture}
  If there is a symmetric tensor field $h_{ab}$ of the second
  rank, whose gauge transformation rule is
  \begin{eqnarray}
    {}_{{\cal Y}}\!h_{ab}
    -
    {}_{{\cal X}}\!h_{ab}
    =
    {\pounds}_{\sigma}g_{ab},
    \label{eq:linear-metric-gauge-trans}
  \end{eqnarray}
  then there exist a tensor field ${\cal H}_{ab}$ and a vector
  field $X^{a}$ such that $h_{ab}$ is decomposed as 
  \begin{eqnarray}
    h_{ab} =: {\cal H}_{ab} + {\pounds}_{X}g_{ab},
    \label{eq:linear-metric-decomp-conjecture}
  \end{eqnarray}
  where ${\cal H}_{ab}$ and $X^{a}$ are transformed as
  \begin{equation}
    {}_{{\cal Y}}\!{\cal H}_{ab} - {}_{{\cal X}}\!{\cal H}_{ab} =  0, 
    \quad
    {}_{\quad{\cal Y}}\!X^{a} - {}_{{\cal X}}\!X^{a} = \sigma^{a}
    \label{eq:linear-metric-decomp-gauge-trans-conjecture}
  \end{equation}
  under the gauge transformation
  (\ref{eq:linear-metric-gauge-trans}), respectively.
\end{conjecture}


In this conjecture, ${\cal H}_{ab}$ is gauge-invariant and we
call ${\cal H}_{ab}$ as {\it gauge-invariant part} of the
perturbation $h_{ab}$.
On the other hand, the vector field $X^{a}$ in equation
(\ref{eq:linear-metric-decomp}) is gauge dependent, and we  
call $X^{a}$ as {\it gauge-variant part} of the perturbation
$h_{ab}$.


In this paper, we assume
Conjecture~\ref{conjecture:decomposition-conjecture}.
This conjecture is quite important in our scenario of the
higher-order gauge-invariant perturbation theory.
In~\cite{kouchan-decomp}, we proposed an outline of a proof of
Conjecture~\ref{conjecture:decomposition-conjecture}.
This outline of a proof is almost complete for an arbitrary
background metric $g_{ab}$.
However, in this outline, there are missing modes for
perturbations, which are called {\it zero modes} and we also
pointed out the physical importance of these zero modes
in~\cite{kouchan-decomp}.
Therefore, we have to say that
Conjecture~\ref{conjecture:decomposition-conjecture} still a
conjecture in our scenario of the higher-order gauge-invariant
perturbation theory.
If we can take these zero modes into our account in the proof of
Conjecture~\ref{conjecture:decomposition-conjecture}, we may
regard that Conjecture~\ref{conjecture:decomposition-conjecture}
is a theorem.


Inspecting the order-by-order gauge-transformation rules
(\ref{eq:Sonego-Bruni-1998-5.1}) and based on
Conjecture~\ref{conjecture:decomposition-conjecture}, we
consider the recursive construction of gauge-invariant variables
for higher-order metric perturbations.
The proposal of this recursive construction is already given in
Sec.~5 of Ref.~\cite{kouchan-gauge-inv}.
In this paper, we try to carry out this proposal through the
gauge-transformation rule (\ref{eq:Sonego-Bruni-1998-5.1}) and
show that this proposal is reduced to
Conjecture~\ref{conjecture:decomposition-conjecture} and
recursive relations of gauge-transformation rules for the 
gauge-variant variables for metric perturbations
(Conjecture~\ref{conjecture:nth-order-gauge-transformation-conjecture}
below).


According to equation (\ref{eq:Sonego-Bruni-1998-5.1}), the
order-by-order gauge-transformation rule for the $n$th-order
metric perturbation ${}_{{\cal X}}^{(n)}\!g_{ab}$ is given by  
\begin{eqnarray}
  {}_{\;{\cal Y}}^{(n)}\!g_{ab}
  -
  {}_{\;{\cal X}}^{(n)}\!g_{ab}
  &=&
  \sum_{l=1}^{n}
  \frac{n!}{(n-l)!}
  \sum_{\{j_{i}\}\in J_{l}} 
  {\cal C}_{l}(\{j_{i}\})
  {\pounds}_{\xi_{(1)}}^{j_{1}}\cdots{\pounds}_{\xi_{(l)}}^{j_{l}}
  {}_{\;\;\;\;\;{\cal X}}^{(n-l)}\!g_{ab}
  .
  \label{eq:Sonego-Bruni-1998-5.1-metric}
\end{eqnarray}
To define the gauge-invariant variables from this
gauge-transformation rule, we reconsider the recursive procedure
to find gauge-invariant variables proposed
in~\cite{kouchan-gauge-inv}.


\subsection{First order}
\label{sec:K.Nakamura-2010-3.1}


Since we assume
Conjecture~\ref{conjecture:decomposition-conjecture} in this
paper and the gauge-transformation rule for the first-order
metric perturbation is given by
\begin{eqnarray}
  {}_{\;{\cal Y}}^{(1)}\!g_{ab}
  -
  {}_{\;{\cal X}}^{(1)}\!g_{ab}
  =
  \sum_{l=1}^{1}
  \frac{1!}{(1-1)!}
  \sum_{\{j_{i}\}\in J_{1}} 
  {\cal C}_{1}(\{j_{i}\})
  {\pounds}_{\xi_{(1)}}^{j_{1}}
  g_{ab}
  =
  {\pounds}_{\xi_{(1)}}g_{ab}
  .
  \label{eq:linear-metric-gauge-trans-0}
\end{eqnarray}
the first-order metric perturbation ${}^{(1)}\!g_{ab}$ is
decomposed as 
\begin{eqnarray}
  &&
  {}^{(1)}\!g_{ab}
  =:
  {}^{(1)}\!{\cal H}_{ab} 
  +
  {\pounds}_{{}^{(1)}\!X}g_{ab}
  ,
  \label{eq:linear-metric-decomp}
  \\
  &&
  {}_{\;{\cal Y}}^{(1)}\!{\cal H}_{ab} 
  -
  {}_{\;{\cal X}}^{(1)}\!{\cal H}_{ab} 
  =
  0
  ,
  \quad
  {}_{\;{\cal Y}}^{(1)}\!X^{a}
  -
  {}_{\;{\cal X}}^{(1)}\!X^{a}
  =
  \xi_{(1)}^{a}
  .
  \label{eq:linear-metric-decomp-gauge-trans}
\end{eqnarray}


Through the gauge-variant vector field ${}^{(1)}\!X^{a}$, we can
define the gauge-invariant variable ${}^{(1)}\!{\cal Q}$ of the
first-order perturbation for an arbitrary tensor field other
than the metric as 
\begin{eqnarray}
  {}^{(1)}\!{\cal Q}
  &:=&
  {}^{(1)}\!Q
  +
  \sum_{l=1}^{1}
  \frac{1!}{(1-l)!}
  \sum_{\{j_{i}\}\in J_{l}}
  {\cal C}_{1}(\{j_{i}\})
  {\pounds}_{-{}^{(1)}\!X}^{j_{1}}
  {}^{(1-l)}\!Q
  \nonumber\\
  &=&
  {}^{(1)}\!Q
  +
  {\pounds}_{-{}^{(1)}\!X}
  {}^{(0)}\!Q
  .
  \label{eq:first-order-gauge-invariant-matter-field}
\end{eqnarray}


\subsection{Second order}
\label{sec:K.Nakamura-2010-3.2}


The gauge-transformation rule for the second-order metric
perturbation is given from equation
(\ref{eq:Sonego-Bruni-1998-5.1-metric}) as 
\begin{eqnarray}
  {}_{\;{\cal Y}}^{(2)}\!g_{ab}
  -
  {}_{\;{\cal X}}^{(2)}\!g_{ab}
  &=&
  \sum_{l=1}^{2}
  \frac{2!}{(2-l)!}
  \sum_{\{j_{i}\}\in J_{l}}
  {\cal C}_{l}(\{j_{i}\})
  {\pounds}_{\xi_{(1)}}^{j_{1}}{\pounds}_{\xi_{(2)}}^{j_{2}}
  {}_{\;\;\;\;\;{\cal X}}^{(2-l)}\!g_{ab}
  \label{eq:2nd-order-gauge-transformation-metric-1}
  \\
  &=&
  2 {\pounds}_{\xi_{(1)}}{}_{\;{\cal X}}^{(1)}\!g_{ab}
  +   \left\{
    {\pounds}_{\xi_{(1)}}^{2} + {\pounds}_{\xi_{(2)}}
  \right\} g_{ab}
  .
  \label{eq:2nd-order-gauge-transformation-metric-2}
\end{eqnarray}
To define the gauge-invariant variables for ${}^{(2)}\!g_{ab}$,
we consider the tensor field defined by 
\begin{eqnarray}
  {}^{(2)}\!\hat{H}_{ab}
  &:=&
  {}^{(2)}\!g_{ab}
  +
  2 {\pounds}_{-{}^{(1)}\!X}{}^{(1)}\!g_{ab}
  + {\pounds}_{-{}^{(1)}\!X}^{2} g_{ab}
  \label{eq:2nd-order-hatHab-def-1}
  \\
  &=&
  {}^{(2)}\!g_{ab}
  +
  \frac{2!}{(2-1)!}
  \sum_{\{j_{i}\}\in J_{1}}
  {\cal C}_{1}(\{j_{i}\})
  {\pounds}_{-{}^{(1)}\!X}^{j_{1}}
  {}^{(1)}\!g_{ab}
  \nonumber\\
  &&
  +
  \frac{2!}{(2-2)!}
  \sum_{\{j_{i}\}\in J_{2}\backslash{}_{2}\!J_{0}^{+}}
  {\cal C}_{2-1}(\{j_{i}\})
  {\pounds}_{-{}^{(1)}\!X}^{j_{1}}
  g_{ab}
  \label{eq:2nd-order-hatHab-def-2}
  ,
\end{eqnarray}
where the vector field ${}^{(1)}\!X^{a}$ is defined as the
gauge-variant part of the first-order metric perturbation
${}^{(1)}\!g_{ab}$ in equation (\ref{eq:linear-metric-decomp})
and ${}_{2}J_{0}^{+}=\{(j_{1},j_{2},...)=(0,1,0,0,...)\}$ is
defined in \ref{sec:K.Nakamura-2014-appendix}.
From the expressions
(\ref{eq:2nd-order-gauge-transformation-metric-1}) and
(\ref{eq:2nd-order-hatHab-def-1}), it is easy to show that the
gauge-transformation rule
\begin{eqnarray}
  {}_{\;{\cal Y}}^{(2)}\!\hat{H}_{ab}
  -
  {}_{\;{\cal X}}^{(2)}\!\hat{H}_{ab}
  =
  {\pounds}_{\sigma_{(2)}}g_{ab}
  ,
  \quad
  \sigma_{(2)}^{a} := \xi_{(2)}^{a} + \hat{\sigma}_{(2)}^{a}
  :=
  \xi_{(2)}^{a} + [\xi_{(1)},{}^{(1)}_{\;{\cal X}}\!X]^{a}
  .
  \label{eq:second-hatHab-gauge-trans-1}
\end{eqnarray}
On the other hand, from the expression
(\ref{eq:2nd-order-hatHab-def-2}), we obtain
\begin{eqnarray}
  &&
  {}_{\;{\cal Y}}^{(2)}\!\hat{H}_{ab}
  -
  {}_{\;{\cal X}}^{(2)}\!\hat{H}_{ab}
  \nonumber\\
  &=&
  {}_{\;{\cal Y}}^{(2)}\!g_{ab}
  -
  {}_{\;{\cal X}}^{(2)}\!g_{ab}
  \nonumber\\
  &&
  +
  \frac{2!}{(2-1)!}
  \sum_{\{j_{i}\}\in J_{1}}
  {\cal C}_{1}(\{j_{i}\})
  \left(
    {\pounds}_{-{}_{\;{\cal Y}}^{(1)}\!X}^{j_{1}}
    {}_{\;{\cal Y}}^{(1)}\!g_{ab}
    -
    {\pounds}_{-{}_{\;{\cal X}}^{(1)}\!X}^{j_{1}}
    {}_{\;{\cal X}}^{(1)}\!g_{ab}
  \right)
  \nonumber\\
  &&
  +
  \frac{2!}{(2-2)!}
  \sum_{\{j_{i}\}\in J_{2}\backslash{}_{2}\!J_{0}^{+}}
  {\cal C}_{2-1}(\{j_{i}\})
  \left(
    {\pounds}_{-{}_{\;{\cal Y}}^{(1)}\!X}^{j_{1}}
    -
    {\pounds}_{-{}_{\;{\cal X}}^{(1)}\!X}^{j_{1}}
  \right)
  g_{ab}
  \nonumber\\
  &=&
  2!
  \sum_{\{j_{i}\}\in J_{1}}
  {\cal C}_{1}(\{j_{i}\})
  \left(
    {\pounds}_{-{}_{\;{\cal Y}}^{(1)}\!X}^{j_{1}}
    -
    {\pounds}_{-{}_{\;{\cal X}}^{(1)}\!X}^{j_{1}}
    +
    {\pounds}_{\xi_{(1)}}^{j_{1}}
  \right)
  {}_{\;{\cal X}}^{(1)}\!g_{ab}
  \nonumber\\
  &&
  +
  2!
  \left[
    \sum_{\{j_{i}\}\in J_{2}\backslash{}_{2}\!J_{0}^{+}}
    {\cal C}_{1}(\{j_{i}\})
    \left(
      {\pounds}_{\xi_{(1)}}^{j_{1}}
      +
      {\pounds}_{-{}_{\;{\cal Y}}^{(1)}\!X}^{j_{1}}
      -
      {\pounds}_{-{}_{\;{\cal X}}^{(1)}\!X}^{j_{1}}
    \right)
  \right.
  \nonumber\\
  &&
  \left.
    +
    \sum_{\{j_{i}\}\in J_{1}}
    {\cal C}_{1}(\{j_{i}\})
    {\pounds}_{-{}_{\;{\cal Y}}^{(1)}\!X}^{j_{1}}
    \sum_{\{k_{m}\}\in J_{1}} 
    {\cal C}_{1}(\{k_{m}\})
    {\pounds}_{\xi_{(1)}}^{k_{1}}
  \right]
  g_{ab}
  \nonumber\\
  &&
  +
  {\pounds}_{\xi_{(2)}}g_{ab}
  .
  \label{eq:second-hatHab-gauge-trans-2}
\end{eqnarray}
Since $J_{1}=\{j_{1}=1,j_{l}=0\;\mbox{for}\;l\geq 2\}$, the
gauge-transformation rule for the variable ${}^{(1)}\!X^{a}$ in
equation (\ref{eq:linear-metric-decomp}) trivially yields 
\begin{eqnarray}
  \label{eq:gauge-trans-identity-first-order}
  \sum_{\{j_{i}\}\in J_{1}}
  {\cal C}_{1}(\{j_{i}\})
  \left(
    {\pounds}_{-{}_{\;{\cal Y}}^{(1)}\!X}^{j_{1}}
    -
    {\pounds}_{-{}_{\;{\cal X}}^{(1)}\!X}^{j_{1}}
    +
    {\pounds}_{\xi_{(1)}}^{j_{1}}
  \right)
  =
  0
  .
\end{eqnarray}
Furthermore, comparing equations
(\ref{eq:second-hatHab-gauge-trans-1}) and
(\ref{eq:second-hatHab-gauge-trans-2}), we obtain the identity
\begin{eqnarray}
  &&
  2!
  \sum_{\{j_{i}\}\in J_{2}\backslash{}_{2}\!J_{0}^{+}}
  {\cal C}_{1}(\{j_{i}\})
  \left(
    {\pounds}_{\xi_{(1)}}^{j_{1}}
    +
    {\pounds}_{-{}_{\;{\cal Y}}^{(1)}\!X}^{j_{1}}
    -
    {\pounds}_{-{}_{\;{\cal X}}^{(1)}\!X}^{j_{1}}
  \right)
  \nonumber\\
  &&
  +
  2!
  \sum_{\{j_{i}\}\in J_{1}}
  {\cal C}_{1}(\{j_{i}\})
  {\pounds}_{-{}_{\;{\cal Y}}^{(1)}\!X}^{j_{1}}
  \sum_{\{k_{m}\}\in J_{1}} 
  {\cal C}_{1}(\{k_{m}\})
  {\pounds}_{\xi_{(1)}}^{k_{1}}
  \nonumber\\
  &=&
  {\pounds}_{\hat{\sigma}_{(2)}}
  .
  \label{eq:gauge-trans-identity-second-order-0}
\end{eqnarray}
Then, we obtain the gauge-transformation rule for the variable
${}^{(2)}\!\hat{H}_{ab}$ as the first equation in equation 
(\ref{eq:second-hatHab-gauge-trans-1}).


Since the gauge-transformation rule for the variable
${}^{(2)}\!\hat{H}_{ab}$ is given in the first equation in
equation (\ref{eq:second-hatHab-gauge-trans-1}), applying
Conjecture~\ref{conjecture:decomposition-conjecture} to the
variable ${}^{(2)}\!\hat{H}_{ab}$, we can decompose
${}^{(2)}\!\hat{H}_{ab}$ as 
\begin{eqnarray}
  {}^{(2)}\!\hat{H}_{ab}
  =:
  {}^{(2)}\!{\cal H}_{ab}
  +
  {\pounds}_{{}^{(2)}\!X}g_{ab},
  \label{eq:second-metric-decomp}
\end{eqnarray}
where the gauge-transformation rules ${}^{(2)}\!{\cal H}_{ab}$
and ${}^{(2)}\!X^{a}$ are given by 
\begin{eqnarray}
  {}_{\;{\cal Y}}^{(2)}\!{\cal H}_{ab}
  -
  {}_{\;{\cal X}}^{(2)}\!{\cal H}_{ab}
  =
  0
  ,
  \quad
  {}_{\;{\cal Y}}^{(2)}\!X^{a}
  -
  {}_{\;{\cal X}}^{(2)}\!X^{a}
  =
  \xi_{(2)}^{a} + \hat{\sigma}_{(2)}^{a}
  .
  \label{eq:second-metric-decomp-gauge-trans}
\end{eqnarray}
Thus, we have decompose the second-order metric perturbation
${}^{(2)}\!g_{ab}$ into its gauge-invariant and gauge-variant
parts as 
\begin{eqnarray}
  {}^{(2)}\!g_{ab}
  =
  {}^{(2)}\!{\cal H}_{ab}
  +
  2 {\pounds}_{{}^{(1)}\!X}{}^{(1)}\!g_{ab}
  +
  \left(
    {\pounds}_{{}^{(2)}\!X}
    - {\pounds}_{{}^{(1)}\!X}^{2}
  \right)g_{ab}
  .
  \label{eq:second-metric-decomp-final}
\end{eqnarray}


The substitution of the second equation in
(\ref{eq:second-metric-decomp-gauge-trans}) into equation
(\ref{eq:gauge-trans-identity-second-order-0}), we obtain
\begin{eqnarray}
  &&
  2!
  \sum_{\{j_{i}\}\in J_{2}\backslash{}_{2}\!J_{0}^{+}}
  {\cal C}_{1}(\{j_{i}\})
  \left(
    {\pounds}_{\xi_{(1)}}^{j_{1}}
    +
    {\pounds}_{-{}_{\;{\cal Y}}^{(1)}\!X}^{j_{1}}
    -
    {\pounds}_{-{}_{\;{\cal X}}^{(1)}\!X}^{j_{1}}
  \right)
  \nonumber\\
  &&
  +
  2!
  \sum_{\{j_{i}\}\in J_{1}}
  {\cal C}_{1}(\{j_{i}\})
  {\pounds}_{-{}_{\;{\cal Y}}^{(1)}\!X}^{j_{1}}
  \sum_{\{k_{m}\}\in J_{1}} 
  {\cal C}_{1}(\{k_{m}\})
  {\pounds}_{\xi_{(1)}}^{k_{1}}
  \nonumber\\
  &=&
  -
  {\pounds}_{\xi_{(2)}}
  -
  {\pounds}_{-{}_{\;{\cal Y}}^{(2)}\!X}
  +
  {\pounds}_{-{}_{\;{\cal X}}^{(2)}\!X}
  .
  \label{eq:gauge-trans-identity-second-order-02}
\end{eqnarray}
It is easy to see that the identity
(\ref{eq:gauge-trans-identity-second-order-02}) is expressed as 
\begin{eqnarray}
  &&
  \sum_{\{j_{i}\}\in J_{2}}
  {\cal C}_{2}(\{j_{i}\})
  \left(
    {\pounds}_{\xi_{(1)}}^{j_{1}}{\pounds}_{\xi_{(2)}}^{j_{2}}
    +
    {\pounds}_{-{}_{\;{\cal Y}}^{(1)}\!X}^{j_{1}}{\pounds}_{-{}_{\;{\cal Y}}^{(2)}\!X}^{j_{2}}
    -
    {\pounds}_{-{}_{\;{\cal X}}^{(1)}\!X}^{j_{1}}{\pounds}_{-{}_{\;{\cal X}}^{(2)}\!X}^{j_{2}}
  \right)
  \nonumber\\
  &&
  +
  \sum_{\{j_{i}\}\in J_{1}}
  {\cal C}_{1}(\{j_{i}\})
  {\pounds}_{-{}_{\;{\cal Y}}^{(1)}\!X}^{j_{1}}
  \sum_{\{k_{m}\}\in J_{1}} 
  {\cal C}_{1}(\{k_{m}\})
  {\pounds}_{\xi_{(1)}}^{k_{1}}
  =
  0
  .
  \label{eq:gauge-trans-identity-second-order-1}
\end{eqnarray}


As shown in~\cite{kouchan-gauge-inv}, through the gauge-variant
variables ${}^{(2)}\!X^{a}$ and ${}^{(1)}\!X^{a}$, we can alway
define the gauge-invariant variables ${}^{(2)}\!{\cal Q}$ for
the second-order perturbation of an arbitrary tensor field other
than the metric as
\begin{eqnarray}
  {}^{(2)}\!{\cal Q}
  &:=&
  {}^{(2)}\!Q
  +
  \sum_{l=1}^{2}
  \frac{2!}{(2-l)!}
  \sum_{\{j_{i}\}\in J_{l}}
  {\cal C}_{l}(\{j_{i}\})
  {\pounds}_{-{}^{(1)}\!X}^{j_{1}}{\pounds}_{-{}^{(2)}\!X}^{j_{l}}
  {}^{(2-l)}\!Q
  \nonumber\\
  &=&
  {}^{(2)}\!Q
  +
  2
  {\pounds}_{-{}^{(1)}\!X}
  {}^{(1)}\!Q
  +
  \left\{
    {\pounds}_{-{}^{(2)}\!X}
    +
    {\pounds}_{-{}^{(1)}\!X}^{2}
  \right\}
  {}^{(0)}\!Q
  .
  \label{eq:second-order-gauge-invariant-matter-field}
\end{eqnarray}


\subsection{Third order}
\label{sec:K.Nakamura-2010-3.3}


The gauge-transformation rule for the third-order metric
perturbation is given from equation
(\ref{eq:Sonego-Bruni-1998-5.1-metric}) as 
\begin{eqnarray}
  {}_{\;{\cal Y}}^{(3)}\!g_{ab}
  -
  {}_{\;{\cal X}}^{(3)}\!g_{ab}
  &=&
  \sum_{l=1}^{3}
  \frac{3!}{(3-l)!}
  \sum_{\{j_{i}\}\in J_{l}} 
  {\cal C}_{l}(\{j_{i}\})
  {\pounds}_{\xi_{(1)}}^{j_{1}}\cdots{\pounds}_{\xi_{(l)}}^{j_{l}}
  {}_{\;\;\;\;\;{\cal X}}^{(3-l)}\!g_{ab}
  \label{eq:3rd-order-gauge-transformation-metric-1}
  \\
  &=&
  3 {\pounds}_{\xi_{(1)}}{}_{\;{\cal X}}^{(2)}\!g_{ab}
  + 3 \left(
    {\pounds}_{\xi_{(1)}}^{2} + {\pounds}_{\xi_{(2)}}
  \right)
  {}_{\;{\cal X}}^{(1)}\!g_{ab}
  \nonumber\\
  &&
  +
  \left(
    {\pounds}_{\xi_{(1)}}^{3}
    + 3 {\pounds}_{\xi_{(1)}}{\pounds}_{\xi_{(2)}}
    + {\pounds}_{\xi_{(3)}}
  \right)
  g_{ab}
  \label{eq:3rd-order-gauge-transformation-metric-2}
  .
\end{eqnarray}
To define the gauge-invariant variables for ${}^{(2)}\!g_{ab}$,
we consider the tensor field defined by 
\begin{eqnarray}
  {}^{(3)}\!\hat{H}_{ab}
  &:=&
  {}^{(3)}\!g_{ab}
  + 3 {\pounds}_{-{}^{(1)}\!X}{}^{(2)}\!g_{ab}
  + 3 \left(
    {\pounds}_{-{}^{(1)}\!X}^{2} + {\pounds}_{-{}^{(2)}\!X}
  \right)
  {}^{(1)}\!g_{ab}
  \nonumber\\
  &&
  +
  \left(
    {\pounds}_{-{}^{(1)}\!X}^{3}
    + 3 {\pounds}_{-{}^{(1)}\!X}{\pounds}_{-{}^{(2)}\!X}
  \right)
  g_{ab}
  \label{eq:3rd-order-hatHab-def-1}
  \\
  &=&
  {}^{(3)}\!g_{ab}
  +
  \sum_{l=1}^{2}
  \frac{3!}{(3-l)!}
  \sum_{\{j_{i}\}\in J_{l}} 
  {\cal C}_{l}(\{j_{i}\})
  {\pounds}_{-{}^{(1)}\!X}^{j_{1}}\cdots{\pounds}_{-{}^{(l)}\!X}^{j_{l}}
  {}^{(3-l)}\!g_{ab}
  \nonumber\\
  &&
  +
  3!
  \sum_{\{j_{i}\}\in J_{3}\backslash{}_{3}\!J_{0}^{+}}
  {\cal C}_{3}(\{j_{i}\})
  {\pounds}_{-{}^{(1)}\!X}^{j_{1}}\cdots{\pounds}_{-{}^{(3)}\!X}^{j_{3}}
  g_{ab}
  \label{eq:3rd-order-hatHab-def-2}
  .
\end{eqnarray}
As shown in \cite{kouchan-gauge-inv}, directly from the
expression (\ref{eq:3rd-order-hatHab-def-1}), we have shown the
gauge-transformation rule for the variable
${}^{(3)}\!\hat{H}_{ab}$ is given as 
\begin{eqnarray}
  &&
  {}_{\;{\cal Y}}^{(3)}\!\hat{H}_{ab}
  -
  {}_{\;{\cal X}}^{(3)}\!\hat{H}_{ab}
  =
  {\pounds}_{\sigma_{(3)}}g_{ab}
  \label{eq:third-hatHab-gauge-trans-1}
  , \\
  &&
  \sigma_{(3)}^{a} := \xi_{(3)}^{a} + \hat{\sigma}_{(3)}^{a}
  , \\
  \label{eq:third-hatHab-gauge-trans-generator-1}
  &&
  \hat{\sigma}_{(3)}^{a}
  :=
  3 [\xi_{(1)},\xi_{(2)}]^{a}
  + 3 [\xi_{(1)},{}_{\;\;{\cal X}}^{(2)}\!X]^{a}
  + 2 [\xi_{(1)},[\xi_{(1)},{}_{\;\;{\cal X}}^{(1)}\!X]]^{a}
  \nonumber\\
  &&
  \quad\quad\quad\quad
  +   [{}_{\;\;{\cal X}}^{(1)}\!X,[\xi_{(1)},{}_{\;\;{\cal X}}^{(1)}\!X]]^{a}
  .
  \label{eq:third-hatHab-gauge-trans-generator-2}
\end{eqnarray}
On the other hand, from the expression
(\ref{eq:3rd-order-hatHab-def-2}), the gauge-transformation rule
for the variable ${}^{(3)}\!\hat{H}_{ab}$ is also given as
\begin{eqnarray}
  &&
  {}_{\;{\cal Y}}^{(3)}\!\hat{H}_{ab}
  -
  {}_{\;{\cal X}}^{(3)}\!\hat{H}_{ab}
  \nonumber\\
  &=&
  \frac{3!}{2!}
  \sum_{\{j_{i}\}\in J_{1}} 
  {\cal C}_{1}(\{j_{i}\})
  \left(
    {\pounds}_{-{}_{\;{\cal Y}}^{(1)}\!X}^{j_{1}}
    -
    {\pounds}_{-{}_{\;{\cal X}}^{(1)}\!X}^{j_{1}}
    +
    {\pounds}_{\xi_{(1)}}^{j_{1}}
  \right)
  {}_{\;{\cal X}}^{(2)}\!g_{ab}
  \nonumber\\
  &&
  +
  3!
  \left[
    \sum_{\{j_{i}\}\in J_{2}} 
    {\cal C}_{2}(\{j_{i}\})
    \left(
      {\pounds}_{-{}_{\;{\cal Y}}^{(1)}\!X}^{j_{1}}
      {\pounds}_{-{}_{\;{\cal Y}}^{(2)}\!X}^{j_{2}}
      -
      {\pounds}_{-{}_{\;{\cal X}}^{(1)}\!X}^{j_{1}}
      {\pounds}_{-{}_{\;{\cal X}}^{(2)}\!X}^{j_{2}}
      -
      {\pounds}_{\xi_{(1)}}^{j_{1}}
      {\pounds}_{\xi_{(2)}}^{j_{2}}
    \right)
  \right.
  \nonumber\\
  && \quad\quad\quad
  \left.
    +
    \sum_{\{j_{i}\}\in J_{1}} 
    {\cal C}_{1}(\{j_{i}\})
    {\pounds}_{-{}_{\;{\cal Y}}^{(1)}\!X}^{j_{1}}
    \sum_{\{k_{m}\}\in J_{1}}
    {\cal C}_{1}(\{k_{m}\})
    {\pounds}_{\xi_{(1)}}^{k_{1}}
  \right]
  {}_{\;{\cal X}}^{(1)}\!g_{ab}
  \nonumber\\
  &&
  +
  3!
  \left[
    \sum_{\{j_{i}\}\in J_{3}\backslash{}_{3}\!J_{0}^{+}}
    {\cal C}_{2}(\{j_{i}\})
    \left(
      {\pounds}_{\xi_{(1)}}^{j_{1}}
      {\pounds}_{\xi_{(2)}}^{j_{2}}
      +
      {\pounds}_{-{}_{\;{\cal Y}}^{(1)}\!X}^{j_{1}}
      {\pounds}_{-{}_{\;{\cal Y}}^{(2)}\!X}^{j_{2}}
      -
      {\pounds}_{-{}_{\;{\cal X}}^{(1)}\!X}^{j_{1}}
      {\pounds}_{-{}_{\;{\cal X}}^{(2)}\!X}^{j_{2}}
    \right)
  \right.
  \nonumber\\
  && \quad\quad\quad
  \left.
    +
    \sum_{\{j_{i}\}\in J_{1}} 
    {\cal C}_{1}(\{j_{i}\})
    {\pounds}_{-{}_{\;{\cal Y}}^{(1)}\!X}^{j_{1}}
    \sum_{\{k_{m}\}\in J_{2}}
    {\cal C}_{2}(\{k_{m}\})
    {\pounds}_{\xi_{(1)}}^{k_{1}}{\pounds}_{\xi_{(2)}}^{k_{2}}
  \right.
  \nonumber\\
  && \quad\quad\quad
  \left.
    +
    \sum_{\{j_{i}\}\in J_{2}} 
    {\cal C}_{2}(\{j_{i}\})
    {\pounds}_{-{}_{\;{\cal Y}}^{(1)}\!X}^{j_{1}}
    {\pounds}_{-{}_{\;{\cal Y}}^{(2)}\!X}^{j_{2}}
    \sum_{\{k_{m}\}\in J_{1}} 
    {\cal C}_{1}(\{k_{m}\})
    {\pounds}_{\xi_{(1)}}^{k_{1}}
  \right]
  g_{ab}
  \nonumber\\
  &&
  +
  {\pounds}_{\xi_{(3)}}
  g_{ab}
  \label{eq:3rd-order-hatHab-gauge-trans-formal-1}
  \\
  &=&
  +
  3!
  \left[
    \sum_{\{j_{i}\}\in J_{3}\backslash{}_{3}\!J_{0}^{+}}
    {\cal C}_{2}(\{j_{i}\})
    \left(
      {\pounds}_{\xi_{(1)}}^{j_{1}}
      {\pounds}_{\xi_{(2)}}^{j_{2}}
      +
      {\pounds}_{-{}_{\;{\cal Y}}^{(1)}\!X}^{j_{1}}
      {\pounds}_{-{}_{\;{\cal Y}}^{(2)}\!X}^{j_{2}}
      -
      {\pounds}_{-{}_{\;{\cal X}}^{(1)}\!X}^{j_{1}}
      {\pounds}_{-{}_{\;{\cal X}}^{(2)}\!X}^{j_{2}}
    \right)
  \right.
  \nonumber\\
  && \quad\quad\quad
  \left.
    +
    \sum_{\{j_{i}\}\in J_{1}} 
    {\cal C}_{1}(\{j_{i}\})
    {\pounds}_{-{}_{\;{\cal Y}}^{(1)}\!X}^{j_{1}}
    \sum_{\{k_{i}\}\in J_{2}}
    {\cal C}_{2}(\{k_{m}\})
    {\pounds}_{\xi_{(1)}}^{k_{1}}{\pounds}_{\xi_{(2)}}^{k_{2}}
  \right.
  \nonumber\\
  && \quad\quad\quad
  \left.
    +
    \sum_{\{j_{i}\}\in J_{2}} 
    {\cal C}_{2}(\{j_{i}\})
    {\pounds}_{-{}_{\;{\cal Y}}^{(1)}\!X}^{j_{1}}
    {\pounds}_{-{}_{\;{\cal Y}}^{(2)}\!X}^{j_{2}}
    \sum_{\{k_{m}\}\in J_{1}} 
    {\cal C}_{1}(\{k_{m}\})
    {\pounds}_{\xi_{(1)}}^{k_{1}}
  \right]
  g_{ab}
  \nonumber\\
  &&
  +
  {\pounds}_{\xi_{(3)}}
  g_{ab}
  \label{eq:3rd-order-hatHab-gauge-trans-formal-2}
  .
\end{eqnarray}
To obtain the expression
(\ref{eq:3rd-order-hatHab-gauge-trans-formal-1}), we used the
lower-order gauge-transformation rules
(\ref{eq:linear-metric-gauge-trans-0}) and
(\ref{eq:2nd-order-gauge-transformation-metric-1}) for the
metric perturbations.
Furthermore, we used the identities
(\ref{eq:gauge-trans-identity-first-order}) and
(\ref{eq:gauge-trans-identity-second-order-1}) to reach the
expression (\ref{eq:3rd-order-hatHab-gauge-trans-formal-2}).


We note that the gauge-transformation rule
(\ref{eq:third-hatHab-gauge-trans-1}) with equation 
(\ref{eq:third-hatHab-gauge-trans-generator-1}) for the
variable ${}^{(3)}\!{\cal H}_{ab}$ yields that 
\begin{eqnarray}
  &&
  3!
  \sum_{\{j_{i}\}\in J_{3}\backslash{}_{3}\!J_{0}^{+}}
  {\cal C}_{2}(\{j_{i}\})
  \left(
    {\pounds}_{\xi_{(1)}}^{j_{1}}
    {\pounds}_{\xi_{(2)}}^{j_{2}}
    +
    {\pounds}_{-{}_{\;{\cal Y}}^{(1)}\!X}^{j_{1}}
    {\pounds}_{-{}_{\;{\cal Y}}^{(2)}\!X}^{j_{2}}
    -
    {\pounds}_{-{}_{\;{\cal X}}^{(1)}\!X}^{j_{1}}
    {\pounds}_{-{}_{\;{\cal X}}^{(2)}\!X}^{j_{2}}
  \right)
  \nonumber\\
  &&
  +
  3!
  \sum_{\{j_{i}\}\in J_{1}} 
  {\cal C}_{1}(\{j_{i}\})
  {\pounds}_{-{}_{\;{\cal Y}}^{(1)}\!X}^{j_{1}}
  \sum_{\{k_{m}\}\in J_{2}}
  {\cal C}_{2}(\{k_{m}\})
  {\pounds}_{\xi_{(1)}}^{k_{1}}{\pounds}_{\xi_{(2)}}^{k_{2}}
  \nonumber\\
  &&
  +
  3!
  \sum_{\{j_{i}\}\in J_{2}} 
  {\cal C}_{2}(\{j_{i}\})
  {\pounds}_{-{}_{\;{\cal Y}}^{(1)}\!X}^{j_{1}}
  {\pounds}_{-{}_{\;{\cal Y}}^{(2)}\!X}^{j_{2}}
  \sum_{\{k_{m}\}\in J_{1}} 
  {\cal C}_{1}(\{k_{m}\})
  {\pounds}_{\xi_{(1)}}^{k_{1}}
  \nonumber\\
  &=&
  {\pounds}_{\hat{\sigma}_{(3)}}
  ,
  \label{eq:gauge-trans-identity-third-order-0}
\end{eqnarray}
since the background metric $g_{ab}$ is arbitrary.


On the other hand, the gauge-transformation rule
(\ref{eq:third-hatHab-gauge-trans-1}) together with
Conjecture~\ref{conjecture:decomposition-conjecture} implies
that the variable ${}^{(3)}\!\hat{H}_{ab}$ is decomposed as 
\begin{eqnarray}
  {}^{(3)}\!\hat{H}_{ab}
  =:
  {}^{(3)}\!{\cal H}_{ab}
  +
  {\pounds}_{{}^{(3)}\!X}g_{ab},
  \label{eq:third-metric-decomp}
\end{eqnarray}
where the gauge-transformation rules ${}^{(3)}\!{\cal H}_{ab}$
and ${}^{(3)}\!X^{a}$ are given by 
\begin{eqnarray}
  {}_{\;{\cal Y}}^{(3)}\!{\cal H}_{ab}
  -
  {}_{\;{\cal X}}^{(3)}\!{\cal H}_{ab}
  =
  0
  ,
  \quad
  {}_{\;{\cal Y}}^{(3)}\!X^{a}
  -
  {}_{\;{\cal X}}^{(3)}\!X^{a}
  =
  \xi_{(3)}^{a} + \hat{\sigma}_{(3)}^{a}
  .
  \label{eq:third-metric-decomp-gauge-trans}
\end{eqnarray}
Thus, we have decompose the third-order metric perturbation
${}^{(3)}\!g_{ab}$ into its gauge-invariant and gauge-variant
parts as 
\begin{eqnarray}
  {}^{(3)}\!g_{ab}
  &:=&
  {}^{(3)}\!{\cal H}_{ab}
  -
  \sum_{l=1}^{3}
  \frac{3!}{(3-l)!}
  \sum_{\{j_{i}\}\in J_{l}} 
  {\cal C}_{l}(\{j_{i}\})
  {\pounds}_{-{}^{(1)}\!X}^{j_{1}}\cdots{\pounds}_{-{}^{(l)}\!X}^{j_{l}}
  {}^{(3-l)}\!g_{ab}
  \label{eq:third-metric-decomp-final-1}
  , \\
  &=&
  {}^{(3)}\!{\cal H}_{ab}
  + 3 {\pounds}_{{}^{(1)}\!X}{}^{(2)}\!g_{ab}
  + 3 \left(
    - {\pounds}_{{}^{(1)}\!X}^{2} + {\pounds}_{{}^{(2)}\!X}
  \right)
  {}^{(1)}\!g_{ab}
  \nonumber\\
  && \quad\quad\quad
  +
  \left(
    {\pounds}_{{}^{(1)}\!X}^{3}
    - 3 {\pounds}_{{}^{(1)}\!X}{\pounds}_{{}^{(2)}\!X}
    +   {\pounds}_{{}^{(3)}\!X}
  \right)
  g_{ab}
  .
  \label{eq:third-metric-decomp-final-2}
\end{eqnarray}


As shown in \cite{kouchan-gauge-inv}, through the gauge-variant
variables ${}^{(1)}\!X^{a}$, ${}^{(2)}\!X^{a}$, and
${}^{(3)}\!X^{a}$, we can always define the gauge-invariant
variables ${}^{(3)}\!{\cal Q}$ for the third-order perturbation 
of an arbitrary tensor field other than the metric as 
\begin{eqnarray}
  {}^{(3)}\!{\cal Q}
  &=&
  {}^{(3)}\!Q
  +
  \sum_{l=1}^{3}
  \frac{3!}{(3-l)!}
  \sum_{\{j_{i}\}\in J_{l}} 
  {\cal C}_{l}(\{j_{i}\})
  {\pounds}_{-{}^{(1)}\!X}^{j_{1}}\cdots{\pounds}_{-{}^{(l)}\!X}^{j_{l}}
  {}^{(3-l)}\!Q
  .
  \label{eq:3rd-order-calQ-def-1}
\end{eqnarray}


Substitution of the second equation in 
(\ref{eq:third-metric-decomp-gauge-trans}) into equation
(\ref{eq:gauge-trans-identity-third-order-0}) leads to the 
identity 
\begin{eqnarray}
  &&
  3!
  \sum_{\{j_{i}\}\in J_{3}\backslash{}_{3}\!J_{0}^{+}}
  {\cal C}_{2}(\{j_{i}\})
  \left(
    {\pounds}_{\xi_{(1)}}^{j_{1}}
    {\pounds}_{\xi_{(2)}}^{j_{2}}
    +
    {\pounds}_{-{}_{\;{\cal Y}}^{(1)}\!X}^{j_{1}}
    {\pounds}_{-{}_{\;{\cal Y}}^{(2)}\!X}^{j_{2}}
    -
    {\pounds}_{-{}_{\;{\cal X}}^{(1)}\!X}^{j_{1}}
    {\pounds}_{-{}_{\;{\cal X}}^{(2)}\!X}^{j_{2}}
  \right)
  \nonumber\\
  &&
  +
  3!
  \sum_{\{j_{i}\}\in J_{1}} 
  {\cal C}_{1}(\{j_{i}\})
  {\pounds}_{-{}_{\;{\cal Y}}^{(1)}\!X}^{j_{1}}
  \sum_{\{k_{i}\}\in J_{2}}
  {\cal C}_{2}(\{k_{m}\})
  {\pounds}_{\xi_{(1)}}^{k_{1}}{\pounds}_{\xi_{(2)}}^{k_{2}}
  \nonumber\\
  &&
  +
  3!
  \sum_{\{j_{i}\}\in J_{2}} 
  {\cal C}_{2}(\{j_{i}\})
  {\pounds}_{-{}_{\;{\cal Y}}^{(1)}\!X}^{j_{1}}
  {\pounds}_{-{}_{\;{\cal Y}}^{(2)}\!X}^{j_{2}}
  \sum_{\{k_{m}\}\in J_{1}} 
  {\cal C}_{1}(\{k_{m}\})
  {\pounds}_{\xi_{(1)}}^{k_{1}}
  \nonumber\\
  &=&
  -
  {\pounds}_{-{}_{\;{\cal Y}}^{(3)}\!X}
  +
  {\pounds}_{-{}_{\;{\cal X}}^{(3)}\!X}
  -
  {\pounds}_{\xi_{(3)}}
  ,
  \label{eq:gauge-trans-identity-third-order-1}
\end{eqnarray}
which is equivalent to the identity 
\begin{eqnarray}
  &&
  \sum_{\{j_{i}\}\in J_{3}}
  {\cal C}_{3}(\{j_{i}\})
  \left(
    {\pounds}_{\xi_{(1)}}^{j_{1}}
    {\pounds}_{\xi_{(2)}}^{j_{2}}
    {\pounds}_{\xi_{(3)}}^{j_{3}}
    +
    {\pounds}_{-{}_{\;{\cal Y}}^{(1)}\!X}^{j_{1}}
    {\pounds}_{-{}_{\;{\cal Y}}^{(2)}\!X}^{j_{2}}
    {\pounds}_{-{}_{\;{\cal Y}}^{(3)}\!X}^{j_{3}}
  \right.
  \nonumber\\
  && \quad\quad\quad\quad\quad\quad\quad\quad\quad\quad\quad
  \left.
    -
    {\pounds}_{-{}_{\;{\cal X}}^{(1)}\!X}^{j_{1}}
    {\pounds}_{-{}_{\;{\cal X}}^{(2)}\!X}^{j_{2}}
    {\pounds}_{-{}_{\;{\cal X}}^{(3)}\!X}^{j_{3}}
  \right)
  \nonumber\\
  &&
  +
  \sum_{\{j_{i}\}\in J_{1}} 
  {\cal C}_{1}(\{j_{i}\})
  {\pounds}_{-{}_{\;{\cal Y}}^{(1)}\!X}^{j_{1}}
  \sum_{\{k_{i}\}\in J_{2}}
  {\cal C}_{2}(\{k_{m}\})
  {\pounds}_{\xi_{(1)}}^{k_{1}}{\pounds}_{\xi_{(2)}}^{k_{2}}
  \nonumber\\
  &&
  +
  \sum_{\{j_{i}\}\in J_{2}} 
  {\cal C}_{2}(\{j_{i}\})
  {\pounds}_{-{}_{\;{\cal Y}}^{(1)}\!X}^{j_{1}}
  {\pounds}_{-{}_{\;{\cal Y}}^{(2)}\!X}^{j_{2}}
  \sum_{\{k_{m}\}\in J_{1}} 
  {\cal C}_{1}(\{k_{m}\})
  {\pounds}_{\xi_{(1)}}^{k_{1}}
  \nonumber\\
  &=&
  0
  .
  \label{eq:gauge-trans-identity-third-order-2}
\end{eqnarray}


\subsection{Fourth order}
\label{sec:K.Nakamura-2010-3.4}


The gauge-transformation rule for the fourth-order metric
perturbation is given from equation
(\ref{eq:Sonego-Bruni-1998-5.1-metric}) as
\begin{eqnarray}
  {}_{\;{\cal Y}}^{(4)}\!g_{ab}
  -
  {}_{\;{\cal X}}^{(4)}\!g_{ab}
  &=&
  \sum_{l=1}^{4}
  \frac{4!}{(4-l)!}
  \sum_{\{j_{i}\}\in J_{l}} 
  {\cal C}_{l}(\{j_{i}\})
  {\pounds}_{\xi_{(1)}}^{j_{1}}\cdots{\pounds}_{\xi_{(l)}}^{j_{l}}
  {}_{\;\;\;\;\;{\cal X}}^{(4-l)}\!g_{ab}
  \label{eq:4th-order-gauge-transformation-metric-1}
  .
\end{eqnarray}
Inspecting this gauge-transformation rule, we define the
gauge-invariant and gauge-variant variables for
${}^{(4)}\!g_{ab}$. 
To do this, as in the case of the second- and third-order
perturbations, we consider the tensor field defined by  
\begin{eqnarray}
  {}^{(4)}\!\hat{H}_{ab}
  &:=&
  {}^{(4)}\!g_{ab}
  +
  \sum_{l=1}^{3}
  \frac{4!}{(4-l)!}
  \sum_{\{j_{i}\}\in J_{l}} 
  {\cal C}_{l}(\{j_{i}\})
  {\pounds}_{-{}^{(1)}\!X}^{j_{1}}\cdots{\pounds}_{-{}^{(l)}\!X}^{j_{l}}
  {}^{(4-l)}\!g_{ab}
  \nonumber\\
  &&
  +
  4!
  \sum_{\{j_{i}\}\in J_{4}\backslash{}_{4}\!J_{0}^{+}} 
  {\cal C}_{3}(\{j_{i}\})
  {\pounds}_{-{}^{(1)}\!X}^{j_{1}}\cdots{\pounds}_{-{}^{(3)}\!X}^{j_{3}}
  g_{ab}
  \label{eq:4th-order-hatHab-def-1}
  ,
\end{eqnarray}
where ${}^{(3)}\!X^{a}$, ${}^{(2)}\!X^{a}$, and
${}^{(1)}\!X^{a}$ are defined previously.
Through the identities
(\ref{eq:gauge-trans-identity-first-order}),
(\ref{eq:gauge-trans-identity-second-order-1}), and
(\ref{eq:gauge-trans-identity-third-order-2}), the
gauge-transformation rule for the variable
${}^{(4)}\!\hat{H}_{ab}$ is given by
\begin{eqnarray}
  &&
  {}_{\;{\cal Y}}^{(4)}\!\hat{H}_{ab}
  -
  {}_{\;{\cal X}}^{(4)}\!\hat{H}_{ab}
  \nonumber\\
  &=&
  {\pounds}_{\xi_{(4)}}g_{ab}
  \nonumber\\
  &&
  +
  4!
  \left[
    \sum_{\{j_{l}\}\in J_{4}\backslash{}_{4}\!J_{0}^{+}}
    {\cal C}_{3}(\{j_{i}\})
    \left(
      {\pounds}_{\xi_{(1)}}^{j_{1}}
      \cdots
      {\pounds}_{\xi_{(3)}}^{j_{3}}
      +
      {\pounds}_{-{}_{\;{\cal Y}}^{(1)}\!X}^{j_{1}}
      \cdots
      {\pounds}_{-{}_{\;{\cal Y}}^{(3)}\!X}^{j_{3}}
    \right.
  \right.
  \nonumber\\
  && \quad\quad\quad\quad\quad\quad\quad\quad\quad\quad\quad\quad\quad\quad
  \left.
    \left.
      -
      {\pounds}_{-{}_{\;{\cal X}}^{(1)}\!X}^{j_{1}}
      \cdots
      {\pounds}_{-{}_{\;{\cal X}}^{(3)}\!X}^{j_{3}}
    \right)
  \right.
  \nonumber\\
  && \quad\quad\quad\quad
  \left.
    +
    \sum_{n=1}^{3}
    \sum_{\{j_{l}\}\in J_{n}}
    {\cal C}_{3}(\{j_{i}\})
    {\pounds}_{-{}_{\;{\cal Y}}^{(1)}\!X}^{j_{1}}
    \cdots
    {\pounds}_{-{}_{\;{\cal Y}}^{(3)}\!X}^{j_{3}}
  \right.
  \nonumber\\
  && \quad\quad\quad\quad\quad\quad\quad
  \left.
    \times
    \sum_{\{k_{m}\}\in J_{4-n}}
    {\cal C}_{3}(\{k_{m}\})
    {\pounds}_{\xi_{(1)}}^{k_{1}}
    \cdots
    {\pounds}_{\xi_{(3)}}^{k_{3}}
  \right]
  g_{ab}
  \label{4th-hatHab-gauge-trans-0}
  .
\end{eqnarray}


Tedious calculations show that the gauge-transformation rule
(\ref{4th-hatHab-gauge-trans-0}) is given by 
\begin{eqnarray}
  {}_{\;{\cal Y}}^{(4)}\!\hat{H}_{ab}
  -
  {}_{\;{\cal X}}^{(4)}\!\hat{H}_{ab}
  =
  {\pounds}_{\sigma_{(4)}}g_{ab}
  ,
  \label{4th-hatHab-gauge-trans-1}
\end{eqnarray}
where $\sigma_{(4)}^{a}$ is given by
\begin{eqnarray}
  \sigma_{(4)}^{a}
  &=&  
  \xi_{(4)}^{a}
  +
  \hat{\sigma}_{(4)}^{a}
  \label{eq:4th-order-sigma-def}
  , \\
  \hat{\sigma}_{(4)}^{a}
  &=&
     4 [\xi_{(1)},\xi_{(3)}]^{a}
  +  6 [\xi_{(1)},[\xi_{(1)},\xi_{(2)}]]^{a}
  +  4 [\xi_{(1)},{}^{(3)}\!X]^{a}
  \nonumber\\
  &&
  +  3 [\xi_{(2)},{}^{(2)}\!X]^{a}
  +  6 [\xi_{(1)},[\xi_{(1)},{}^{(2)}\!X]]^{a}
  +  3 [\xi_{(2)},[\xi_{(1)},{}^{(1)}\!X]]^{a}
  \nonumber\\
  &&
  +  3 [{}^{(2)}\!X,[\xi_{(1)},{}^{(1)}\!X]]^{a}
  +  3 [\xi_{(1)},[\xi_{(1)},[\xi_{(1)},{}^{(1)}\!X]]]^{a}
  \nonumber\\
  &&
  +  3 [\xi_{(1)},[{}^{(1)}\!X,[\xi_{(1)},{}^{(1)}\!X]]]^{a}
  +    [{}^{(1)}\!X,[{}^{(1)}\!X,[\xi_{(1)},{}^{(1)}\!X]]]^{a}
  .
  \label{eq:4th-order-hatsigma-def}
\end{eqnarray}
Then, we may apply
Conjecture~\ref{conjecture:decomposition-conjecture} to the
variable ${}^{(4)}\!\hat{H}_{ab}$, we can decompose
${}^{(4)}\!\hat{H}_{ab}$ into its gauge-invariant and 
gauge-variant parts as 
\begin{eqnarray}
  {}^{(4)}\!\hat{H}_{ab}
  =:
  {}^{(4)}\!{\cal H}_{ab}
  +
  {\pounds}_{{}^{(4)}\!X}g_{ab}
  ,
  \label{eq:fourth-metric-decomp}
\end{eqnarray}
where the gauge-transformation rules for the variables 
${}^{(4)}\!{\cal H}_{ab}$ and ${}^{(4)}\!X^{a}$ is given by 
\begin{eqnarray}
  &&
  {}_{\;\;{\cal Y}}^{(4)}\!{\cal H}_{ab}
  -
  {}_{\;\;{\cal X}}^{(4)}\!{\cal H}_{ab}
  =
  0
  ,
  \quad
  {}_{\;\;{\cal Y}}^{(4)}\!X^{a}
  -
  {}_{\;\;{\cal X}}^{(4)}\!X^{a}
  =
  \sigma_{(4)}^{a}
  =
  \xi_{(4)}^{a} + \hat{\sigma}_{(4)}^{a}
  .
  \label{eq:fourth-metric-decomp-gauge-trans}
\end{eqnarray}
Thus, we have decompose the fourth-order metric perturbation
${}^{(4)}\!g_{ab}$ into its gauge-invariant and gauge-variant
parts as 
\begin{eqnarray}
  {}^{(4)}\!g_{ab}
  &=&
  {}^{(4)}\!{\cal H}_{ab}
  -
  \sum_{l=1}^{4}
  \frac{4!}{(4-l)!}
  \sum_{\{j_{i}\}\in J_{l}} 
  {\cal C}_{l}(\{j_{i}\})
  {\pounds}_{-{}^{(1)}\!X}^{j_{1}}\cdots{\pounds}_{-{}^{(l)}\!X}^{j_{l}}
  {}^{(4-l)}\!g_{ab}
  \label{eq:fourth-metric-decomp-final-1}
  .
\end{eqnarray}


As in the case of the lower-order perturbations, we can always
define the gauge-invariant variables ${}^{(4)}\!{\cal Q}$ for
the fourth-order perturbation of an arbitrary tensor field other
than the metric through the gauge-variant parts
${}^{(1)}\!X^{a}$, ${}^{(2)}\!X^{a}$, ${}^{(3)}\!X^{a}$, and
${}^{(4)}\!X^{a}$ of the metric perturbations:
\begin{eqnarray}
  {}^{(4)}\!{\cal Q}
  &:=&
  {}^{(4)}\!Q
  +
  \sum_{l=1}^{4}
  \frac{4!}{(4-l)!}
  \sum_{\{j_{i}\}\in J_{l}} 
  {\cal C}_{l}(\{j_{i}\})
  {\pounds}_{-{}^{(1)}\!X}^{j_{1}}\cdots{\pounds}_{-{}^{(l)}\!X}^{j_{l}}
  {}^{(4-l)}\!Q
  .
  \label{eq:4th-order-calQ-def-1}
\end{eqnarray}


We also note that the gauge-transformation rules
(\ref{4th-hatHab-gauge-trans-0}),
(\ref{4th-hatHab-gauge-trans-1}), and the second equation in
(\ref{eq:fourth-metric-decomp-gauge-trans}) implies
the identity 
\begin{eqnarray}
  &&
  4!
  \sum_{\{j_{l}\}\in J_{4}\backslash{}_{4}\!J_{0}^{+}}
  {\cal C}_{3}(\{j_{i}\})
  \left(
    {\pounds}_{\xi_{(1)}}^{j_{1}}
    \cdots
    {\pounds}_{\xi_{(3)}}^{j_{3}}
    +
    {\pounds}_{-{}_{\;{\cal Y}}^{(1)}\!X}^{j_{1}}
    \cdots
    {\pounds}_{-{}_{\;{\cal Y}}^{(3)}\!X}^{j_{3}}
  \right.
  \nonumber\\
  && \quad\quad\quad\quad\quad\quad\quad\quad\quad\quad\quad\quad
  \left.
    -
    {\pounds}_{-{}_{\;{\cal X}}^{(1)}\!X}^{j_{1}}
    \cdots
    {\pounds}_{-{}_{\;{\cal X}}^{(3)}\!X}^{j_{3}}
  \right)
  \nonumber\\
  && \quad\quad\quad\quad
  +
  4!
  \sum_{n=1}^{3}
  \sum_{\{j_{l}\}\in J_{n}}
  {\cal C}_{3}(\{j_{i}\})
  {\pounds}_{-{}_{\;{\cal Y}}^{(1)}\!X}^{j_{1}}
  \cdots
  {\pounds}_{-{}_{\;{\cal Y}}^{(3)}\!X}^{j_{3}}
  \nonumber\\
  && \quad\quad\quad\quad\quad\quad\quad
  \times
  \sum_{\{k_{m}\}\in J_{4-n}}
  {\cal C}_{3}(\{k_{m}\})
  {\pounds}_{\xi_{(1)}}^{k_{1}}
  \cdots
  {\pounds}_{\xi_{(3)}}^{k_{3}}
  \nonumber\\
  &=&
  {\pounds}_{\hat{\sigma}_{(4)}}
  .
  \label{4th-hatHab-gauge-trans-2}
\end{eqnarray}
Substituting the second equation in
(\ref{eq:fourth-metric-decomp-gauge-trans}) into
(\ref{4th-hatHab-gauge-trans-2}), we obtain the identity 
\begin{eqnarray}
  &&
  4!
  \sum_{\{j_{l}\}\in J_{4}\backslash{}_{4}\!J_{0}^{+}}
  {\cal C}_{3}(\{j_{i}\})
  \left(
    {\pounds}_{\xi_{(1)}}^{j_{1}}
    \cdots
    {\pounds}_{\xi_{(3)}}^{j_{3}}
    +
    {\pounds}_{-{}_{\;{\cal Y}}^{(1)}\!X}^{j_{1}}
    \cdots
    {\pounds}_{-{}_{\;{\cal Y}}^{(3)}\!X}^{j_{3}}
  \right.
  \nonumber\\
  && \quad\quad\quad\quad\quad\quad\quad\quad\quad\quad\quad\quad
  \left.
      -
      {\pounds}_{-{}_{\;{\cal X}}^{(1)}\!X}^{j_{1}}
      \cdots
      {\pounds}_{-{}_{\;{\cal X}}^{(3)}\!X}^{j_{3}}
    \right)
  \nonumber\\
  && \quad\quad\quad\quad
  +
  4!
  \sum_{n=1}^{3}
  \sum_{\{j_{i}\}\in J_{n}}
  {\cal C}_{3}(\{j_{i}\})
  {\pounds}_{-{}_{\;{\cal Y}}^{(1)}\!X}^{j_{1}}
  \cdots
  {\pounds}_{-{}_{\;{\cal Y}}^{(3)}\!X}^{j_{3}}
  \nonumber\\
  && \quad\quad\quad\quad\quad\quad\quad
  \times
  \sum_{\{k_{m}\}\in J_{4-n}}
  {\cal C}_{3}(\{k_{m}\})
  {\pounds}_{\xi_{(1)}}^{k_{1}}
  \cdots
  {\pounds}_{\xi_{(3)}}^{k_{3}}
  \nonumber\\
  &=&
  -
  {\pounds}_{\xi_{(4)}}
  -
  {\pounds}_{-{}_{\;{\cal Y}}^{(4)}\!X}
  +
  {\pounds}_{-{}_{\;{\cal X}}^{(4)}\!X}
  .
  \label{eq:gauge-trans-identity-fourth-order-1}
\end{eqnarray}
This identity is also expressed as 
\begin{eqnarray}
  &&
  \sum_{\{j_{l}\}\in J_{4}}
  {\cal C}_{4}(\{j_{i}\})
  \left(
    {\pounds}_{\xi_{(1)}}^{j_{1}}
    \cdots
    {\pounds}_{\xi_{(3)}}^{j_{4}}
    +
    {\pounds}_{-{}_{\;{\cal Y}}^{(1)}\!X}^{j_{1}}
    \cdots
    {\pounds}_{-{}_{\;{\cal Y}}^{(3)}\!X}^{j_{4}}
  \right.
  \nonumber\\
  && \quad\quad\quad\quad\quad\quad\quad\quad\quad\quad\quad\quad
  \left.
      -
      {\pounds}_{-{}_{\;{\cal X}}^{(1)}\!X}^{j_{1}}
      \cdots
      {\pounds}_{-{}_{\;{\cal X}}^{(4)}\!X}^{j_{4}}
    \right)
  \nonumber\\
  && 
  +
  \sum_{n=1}^{3}
  \sum_{\{j_{l}\}\in J_{n}}
  {\cal C}_{3}(\{j_{i}\})
  {\pounds}_{-{}_{\;{\cal Y}}^{(1)}\!X}^{j_{1}}
  \cdots
  {\pounds}_{-{}_{\;{\cal Y}}^{(3)}\!X}^{j_{3}}
  \nonumber\\
  && \quad\quad\quad\quad
  \times
  \sum_{\{k_{m}\}\in J_{4-n}}
  {\cal C}_{3}(\{k_{m}\})
  {\pounds}_{\xi_{(1)}}^{k_{1}}
  \cdots
  {\pounds}_{\xi_{(3)}}^{k_{3}}
  \nonumber\\
  &=&
  0
  ,
  \label{eq:gauge-trans-identity-fourth-order-2}
\end{eqnarray}
or, equivalently, 
\begin{eqnarray}
  &&
  \sum_{n=1}^{4}
  \sum_{\{j_{l}\}\in J_{n}}
  {\cal C}_{4}(\{j_{i}\})
  {\pounds}_{-{}_{\;{\cal Y}}^{(1)}\!X}^{j_{1}}
  \cdots
  {\pounds}_{-{}_{\;{\cal Y}}^{(4)}\!X}^{j_{4}}
  \sum_{\{k_{m}\}\in J_{4-n}}
  {\cal C}_{4}(\{k_{m}\})
  {\pounds}_{\xi_{(1)}}^{k_{1}}
  \cdots
  {\pounds}_{\xi_{(4)}}^{k_{4}}
  \nonumber\\
  &=&
  \sum_{\{j_{l}\}\in J_{4}}
  {\cal C}_{4}(\{j_{i}\})
  {\pounds}_{-{}_{\;{\cal X}}^{(1)}\!X}^{j_{1}}
  \cdots
  {\pounds}_{-{}_{\;{\cal X}}^{(4)}\!X}^{j_{4}}
  .
  \label{eq:gauge-trans-identity-fourth-order-3}
\end{eqnarray}


\section{Recursive structure in the definitions of
  gauge-invariant variables for $n$th-order perturbations}
\label{sec:K.Nakamura-2010-4}


In the last section, we have shown the construction of
gauge-invariant variables to 4th order.
From these construction, we easily expect that it can be
generalize to $n$th-order perturbations.
In this section, we show the scenario of the generalization of
the construction of gauge-invariant variables to $n$th order
which can be expected from the results in the last section.


As noted in section~\ref{sec:K.Nakamura-2014-3}, the
gauge-transformation rule for the $n$th-order metric
perturbation is given by equation
(\ref{eq:Sonego-Bruni-1998-5.1-metric}).
Inspecting this gauge-transformation rule, we construct the
gauge-invariant variables for ${}^{(n)}\!g_{ab}$.
Through the construction of gauge-invariant variables for
${}^{(i)}\!g_{ab}$ ($i=1,...,n-1$), we can also define the
vector fields ${}^{(i)}\!X^{a}$ ($i=1,...,n-1$) are defined
through the construction.
\begin{eqnarray}
  {}^{(i)}_{\;{\cal Y}}\!X^{a}
  -
  {}^{(i)}_{\;{\cal X}}\!X^{a}
  =
  \sigma_{(i)}^{a}
  =
  \xi_{(i)}^{a} + \hat{\sigma}_{(i)}^{a}
  .
  \label{eq:ith-metric-decomp-gauge-trans}
\end{eqnarray}
Furthermore, we can also obtain the $n-1$ identities which are
expressed as 
\begin{eqnarray}
  &&
  \sum_{p=1}^{i}
  \sum_{\{j_{l}\}\in J_{p}}
  {\cal C}_{i}(\{j_{l}\})
  {\pounds}_{-{}_{\;{\cal Y}}^{(1)}\!X}^{j_{1}}
  \cdots
  {\pounds}_{-{}_{\;{\cal Y}}^{(i)}\!X}^{j_{i}}
  \sum_{\{k_{m}\}\in J_{i-p}}
  {\cal C}_{i}(\{k_{m}\})
  {\pounds}_{\xi_{(1)}}^{k_{1}}
  \cdots
  {\pounds}_{\xi_{(i)}}^{k_{i}}
  \nonumber\\
  &=&
  \sum_{\{j_{l}\}\in J_{i}}
  {\cal C}_{i}(\{j_{l}\})
  {\pounds}_{-{}_{\;{\cal X}}^{(1)}\!X}^{j_{1}}
  \cdots
  {\pounds}_{-{}_{\;{\cal X}}^{(4)}\!X}^{j_{i}}
  .
  \label{eq:gauge-trans-identity-n-1}
\end{eqnarray}


To define construct the gauge-invariant variables for the metric
perturbation ${}^{(n)}\!g_{ab}$, as in the cases in the last
section, we consider the tensor field defined by
\begin{eqnarray}
  {}^{(n)}\!\hat{H}_{ab}
  &:=&
  {}^{(n)}\!g_{ab}
  +
  \sum_{l=1}^{n-1}
  \frac{n!}{(n-l)!}
  \sum_{\{j_{i}\}\in J_{l}} 
  {\cal C}_{l}(\{j_{i}\})
  {\pounds}_{-{}^{(1)}\!X}^{j_{1}}\cdots{\pounds}_{-{}^{(l)}\!X}^{j_{l}}
  {}^{(n-l)}\!g_{ab}
  \nonumber\\
  &&
  +
  n!
  \sum_{\{j_{i}\}\in J_{n}\backslash{}_{n}\!J_{0}^{+}} 
  {\cal C}_{n-1}(\{j_{i}\})
  {\pounds}_{-{}^{(1)}\!X}^{j_{1}}\cdots{\pounds}_{-{}^{(n-1)}\!X}^{j_{n-1}}
  g_{ab}
  \label{eq:nth-order-hatHab-def}
  .
\end{eqnarray}
Using the order-by-order identities
(\ref{eq:gauge-trans-identity-n-1}), the
gauge-transformation rule is given by 
\begin{eqnarray}
  &&
  {}_{\;{\cal Y}}^{(n)}\!\hat{H}_{ab}
  -
  {}_{\;{\cal X}}^{(n)}\!\hat{H}_{ab}
  \nonumber\\
  &=&
  {\pounds}_{\xi_{(n)}}g_{ab}
  \nonumber\\
  &&
  +
  n!
  \left[
    \sum_{\{j_{l}\}\in J_{n}\backslash{}_{n}\!J_{0}^{+}}
    {\cal C}_{n-1}(\{j_{l}\})
    \left(
      {\pounds}_{\xi_{(1)}}^{j_{1}}
      \cdots
      {\pounds}_{\xi_{(n-1)}}^{j_{n-1}}
      +
      {\pounds}_{-{}_{\;{\cal Y}}^{(1)}\!X}^{j_{1}}
      \cdots
      {\pounds}_{-{}_{\;\;\;\;\;\;{\cal Y}}^{(n-1)}\!X}^{j_{n-1}}
    \right.
  \right.
  \nonumber\\
  && \quad\quad\quad\quad\quad\quad\quad\quad\quad\quad\quad\quad\quad\quad
  \left.
    \left.
      -
      {\pounds}_{-{}_{\;{\cal X}}^{(1)}\!X}^{j_{1}}
      \cdots
      {\pounds}_{-{}_{\;\;\;\;\;\;{\cal X}}^{(n-1)}\!X}^{j_{n-1}}
    \right)
  \right.
  \nonumber\\
  && \quad\quad\quad\quad
  \left.
    +
    \sum_{i=1}^{n-1}
    \sum_{\{j_{l}\}\in J_{i}}
    {\cal C}_{n-1}(\{j_{l}\})
    {\pounds}_{-{}_{\;{\cal Y}}^{(1)}\!X}^{j_{1}}
    \cdots
    {\pounds}_{-{}_{\;\;\;\;\;\;{\cal Y}}^{(3)}\!X}^{j_{n-1}}
  \right.
  \nonumber\\
  && \quad\quad\quad\quad\quad\quad\quad
  \left.
    \times
    \sum_{\{k_{m}\}\in J_{n-i}}
    {\cal C}_{n-1}(\{k_{m}\})
    {\pounds}_{\xi_{(1)}}^{k_{1}}
    \cdots
    {\pounds}_{\xi_{(n-1)}}^{k_{n-1}}
  \right]
  g_{ab}
  \label{nth-hatHab-gauge-trans-0}
  .
\end{eqnarray}
From the analyses in the last section, we can expect that the 
following conjecture is reasonable.


\begin{conjecture}
  \label{conjecture:nth-order-gauge-transformation-conjecture}
  There exists a vector field $\hat{\sigma}_{(n)}^{a}$ such that 
  \begin{eqnarray}
    &&
    n!
    \sum_{\{j_{l}\}\in J_{n}\backslash{}_{n}\!J_{0}^{+}}
    {\cal C}_{n-1}(\{j_{l}\})
    \left(
      {\pounds}_{\xi_{(1)}}^{j_{1}}
      \cdots
      {\pounds}_{\xi_{(n-1)}}^{j_{n-1}}
      +
      {\pounds}_{-{}_{\;{\cal Y}}^{(1)}\!X}^{j_{1}}
      \cdots
      {\pounds}_{-{}_{\;\;\;\;\;\;{\cal Y}}^{(n-1)}\!X}^{j_{n-1}}
    \right.
    \nonumber\\
    && \quad\quad\quad\quad\quad\quad\quad\quad\quad\quad\quad\quad
    \left.
      -
      {\pounds}_{-{}_{\;{\cal X}}^{(1)}\!X}^{j_{1}}
      \cdots
      {\pounds}_{-{}_{\;\;\;\;\;\;{\cal X}}^{(n-1)}\!X}^{j_{n-1}}
    \right)
    \nonumber\\
    && \quad\quad\quad\quad
    +
    n!
    \sum_{i=1}^{n-1}
    \sum_{\{j_{l}\}\in J_{i}}
    {\cal C}_{n-1}(\{j_{l}\})
    {\pounds}_{-{}_{\;{\cal Y}}^{(1)}\!X}^{j_{1}}
    \cdots
    {\pounds}_{-{}_{\;\;\;\;\;\;{\cal Y}}^{(n-1)}\!X}^{j_{n-1}}
    \nonumber\\
    && \quad\quad\quad\quad\quad\quad\quad
    \times
    \sum_{\{k_{m}\}\in J_{n-i}}
    {\cal C}_{n-1}(\{k_{m}\})
    {\pounds}_{\xi_{(1)}}^{k_{1}}
    \cdots
    {\pounds}_{\xi_{(n-1)}}^{k_{n-1}}
    \nonumber\\
    &=&
    {\pounds}_{\hat{\sigma}_{(n)}}
    .
    \label{nth-hatHab-conjectured-identity-1}
  \end{eqnarray}
\end{conjecture}
To derive the explicit form of $\hat{\sigma}_{(n)}$, tough
algebraic calculations are necessary.
Although we do not going to the details of the proof of this
conjecture, we expect that this identity should be proved,
recursively, and there will be no difficulty except for tough
algebraic calculations. 
Actually, in the last section, we have confirmed this conjecture 
to 4th order and it is reasonable to regard that
Conjecture~\ref{conjecture:nth-order-gauge-transformation-conjecture}
is hold.


If Conjecture 
\ref{conjecture:nth-order-gauge-transformation-conjecture} is
hold, the gauge-transformation rule for the variable
${}^{(n)}\!H_{ab}$ is given by
\begin{eqnarray}
  {}_{\;{\cal Y}}^{(n)}\!\hat{H}_{ab}
  -
  {}_{\;{\cal X}}^{(n)}\!\hat{H}_{ab}
  =
  {\pounds}_{\sigma_{(n)}}g_{ab}
  , \quad
  \sigma_{(n)}^{a} := \xi_{(n)}^{a} + \hat{\sigma}_{(n)}^{a}.
  \label{eq:nth-hatHab-gauge-trans-final}
\end{eqnarray}
This is the same form as the gauge-transformation rule for the
linear-order metric perturbation.
Then, we may apply
Conjecture~\ref{conjecture:decomposition-conjecture} for the
variable ${}^{(n)}\!\hat{H}_{ab}$.
This implies that the variable ${}^{(n)}\!\hat{H}_{ab}$ is
decomposed as
\begin{eqnarray}
  &&
  {}^{(n)}\!\hat{H}_{ab}
  =
  {}^{(n)}\!{\cal H}_{ab}
  +
  {\pounds}_{{}^{(n)}\!X}g_{ab}
  \label{eq:nth-hatHab-decomp}
  ,\\
  &&
  {}_{\;{\cal Y}}^{(n)}\!{\cal H}_{ab}
  -
  {}_{\;{\cal X}}^{(n)}\!{\cal H}_{ab}
  =
  0
  ,
  \quad
  {}_{\;\;{\cal Y}}^{(n)}\!X^{a}
  -
  {}_{\;\;{\cal X}}^{(n)}\!X^{a}
  =
  \sigma_{(n)}^{a}
  =
  \xi_{(n)}^{a} + \hat{\sigma}_{(n)}^{a}
  .
  \label{eq:nth-metric-decomp-gauge-trans}
\end{eqnarray}
Thus, we have gauge-invariant variables 
${}^{(n)}\!{\cal H}_{ab}$ for the $n$th-order metric
perturbation. 
This implies that the original $n$th-order metric perturbation
${}^{(n)}\!g_{ab}$ 
\begin{eqnarray}
  {}^{(n)}\!g_{ab}
  &=&
  {}^{(n)}\!{\cal H}_{ab}
  -
  {\pounds}_{-{}^{(n)}\!X}g_{ab}
  \nonumber\\
  &&
  -
  \sum_{l=1}^{n-1}
  \frac{n!}{(n-l)!}
  \sum_{\{j_{i}\}\in J_{l}} 
  {\cal C}_{l}(\{j_{i}\})
  {\pounds}_{-{}^{(1)}\!X}^{j_{1}}\cdots{\pounds}_{-{}^{(l)}\!X}^{j_{l}}
  {}^{(n-l)}\!g_{ab}
  \nonumber\\
  &&
  -
  n!
  \sum_{\{j_{i}\}\in J_{n}\backslash{}_{n}\!J_{0}^{+}} 
  {\cal C}_{n-1}(\{j_{i}\})
  {\pounds}_{-{}^{(1)}\!X}^{j_{1}}\cdots{\pounds}_{-{}^{(n-1)}\!X}^{j_{n-1}}
  g_{ab}
  \nonumber\\
  &=&
  {}^{(n)}\!{\cal H}_{ab}
  -
  \sum_{l=1}^{n}
  \frac{n!}{(n-l)!}
  \sum_{\{j_{i}\}\in J_{l}} 
  {\cal C}_{l}(\{j_{i}\})
  {\pounds}_{-{}^{(1)}\!X}^{j_{1}}\cdots{\pounds}_{-{}^{(l)}\!X}^{j_{l}}
  {}^{(n-l)}\!g_{ab}
  .
  \label{eq:nth-order-original-ngab-decomp}
\end{eqnarray}
This indicate that the $n$th-order metric perturbation
${}^{(n)}\!g_{ab}$ is decomposed as its gauge-invariant, and
gauge-variant parts.
Through the gauge-variant variables ${}^{(i)}\!X^{a}$
($i=1,...,n$), we can also define the gauge-invariant variable
${}^{(n)}\!{\cal Q}$ for the $n$th-order perturbation
${}^{(n)}\!Q$ of any tensor field $Q$ is also defined as
\begin{eqnarray}
  {}^{(n)}\!{\cal Q}
  &:=&
  {}^{(n)}\!Q
  +
  \sum_{l=1}^{n}
  \frac{n!}{(n-l)!}
  \sum_{\{j_{i}\}\in J_{l}} 
  {\cal C}_{l}(\{j_{i}\})
  {\pounds}_{-{}^{(1)}\!X}^{j_{1}}\cdots{\pounds}_{-{}^{(l)}\!X}^{j_{l}}
  {}^{(n-l)}\!Q
  .
  \label{eq:nth-order-calQ-def}
\end{eqnarray}


Furthermore,
Conjecture~\ref{conjecture:nth-order-gauge-transformation-conjecture}
leads the identity which corresponds to
(\ref{eq:gauge-trans-identity-first-order}),
(\ref{eq:gauge-trans-identity-second-order-1}),
(\ref{eq:gauge-trans-identity-third-order-2}), and
(\ref{eq:gauge-trans-identity-fourth-order-3}).
Substituting the second equation in
(\ref{eq:nth-metric-decomp-gauge-trans}) into equation
(\ref{nth-hatHab-conjectured-identity-1}), we obtain
\begin{eqnarray}
  &&
  n!
  \sum_{\{j_{l}\}\in J_{n}\backslash{}_{n}\!J_{0}^{+}}
  {\cal C}_{n-1}(\{j_{l}\})
  \left(
    {\pounds}_{\xi_{(1)}}^{j_{1}}
    \cdots
    {\pounds}_{\xi_{(n-1)}}^{j_{n-1}}
    +
    {\pounds}_{-{}_{\;{\cal Y}}^{(1)}\!X}^{j_{1}}
    \cdots
    {\pounds}_{-{}_{\;\;\;\;\;\;{\cal Y}}^{(n-1)}\!X}^{j_{n-1}}
  \right.
  \nonumber\\
  && \quad\quad\quad\quad\quad\quad\quad\quad\quad\quad\quad\quad
  \left.
    -
    {\pounds}_{-{}_{\;{\cal X}}^{(1)}\!X}^{j_{1}}
    \cdots
    {\pounds}_{-{}_{\;\;\;\;\;\;{\cal X}}^{(n-1)}\!X}^{j_{n-1}}
  \right)
  \nonumber\\
  && \quad\quad\quad\quad
  +
  n!
  \sum_{i=1}^{n-1}
  \sum_{\{j_{l}\}\in J_{i}}
  {\cal C}_{n-1}(\{j_{l}\})
  {\pounds}_{-{}_{\;{\cal Y}}^{(1)}\!X}^{j_{1}}
  \cdots
  {\pounds}_{-{}_{\;\;\;\;\;\;{\cal Y}}^{(n-1)}\!X}^{j_{n-1}}
  \nonumber\\
  && \quad\quad\quad\quad\quad\quad\quad
  \times
  \sum_{\{k_{m}\}\in J_{n-i}}
  {\cal C}_{n-1}(\{k_{m}\})
  {\pounds}_{\xi_{(1)}}^{k_{1}}
  \cdots
  {\pounds}_{\xi_{(n-1)}}^{k_{n-1}}
  \nonumber\\
  &=&
  -
  {\pounds}_{\xi_{(n)}}
  -
  {\pounds}_{-{}_{\;\;{\cal Y}}^{(n)}\!X}
  +
  {\pounds}_{-{}_{\;\;{\cal X}}^{(n)}\!X}
  .
  \label{nth-hatHab-conjectured-identity-2}
\end{eqnarray}
This is equivalent to
\begin{eqnarray}
  &&
  \sum_{i=1}^{n}
  \sum_{\{j_{l}\}\in J_{i}}
  {\cal C}_{n}(\{j_{l}\})
  {\pounds}_{-{}_{\;{\cal Y}}^{(1)}\!X}^{j_{1}}
  \cdots
  {\pounds}_{-{}_{\;{\cal Y}}^{(n)}\!X}^{j_{n}}
  \sum_{\{k_{m}\}\in J_{n-i}}
  {\cal C}_{n}(\{k_{m}\})
  {\pounds}_{\xi_{(1)}}^{k_{1}}
  \cdots
  {\pounds}_{\xi_{(n)}}^{k_{n}}
  \nonumber\\
  &=&
  \sum_{\{j_{l}\}\in J_{n}}
  {\cal C}_{n}(\{j_{i}\})
  {\pounds}_{-{}_{\;{\cal X}}^{(1)}\!X}^{j_{1}}
  \cdots
  {\pounds}_{-{}_{\;{\cal X}}^{(n)}\!X}^{j_{n}}
  .
  \label{eq:gauge-trans-identity-fourth-order-n}
\end{eqnarray}
This identity corresponds to the $i=n$ version of identities
(\ref{eq:gauge-trans-identity-n-1}) and is used when we derive
the gauge-transformation rules of perturbations higher than
$n$th.


\section{Example: Cosmological Perturbations}
\label{sec:K.Nakamura-2010-5}


Here, we consider the application of our formulae derived in the 
last section to a specific background spacetime as an example.
The example discussed here is the cosmological perturbation
whose background metric is given by 
\begin{eqnarray}
  \label{eq:cosmological-background-metric}
  g_{ab}
  =
  a^{2}(\eta)\left(
    - (d\eta)_{a}(d\eta)_{b}
    + \gamma_{pq}(dx^{p})_{a}(dx^{q})_{b}
  \right)
  ,
\end{eqnarray}
where $a=a(\eta)$ is the scale factor, $\gamma_{pq}$ is the
metric on the maximally symmetric 3-space with curvature
constant $K$, and the indices $p,q,r,...$ for the spatial
components run from 1 to 3.
In this section, we concentrate only on the metric
perturbations.


We have to note that even in the case of this cosmological
perturbations, there is the ``zero-mode problem'' which is
mentioned in Sec.~\ref{sec:K.Nakamura-2014-3}.
In this section, we ignore these zero-modes and assume
Conjecture~\ref{conjecture:decomposition-conjecture}, for 
simplicity, because we have not yet resolved the ``zero-mode
problem'' systematically as mentioned in
Sec.~\ref{sec:K.Nakamura-2014-3}.


On the background spacetime with the metric
(\ref{eq:cosmological-background-metric}), we consider the
metric perturbation ${}^{(1)}\!g_{ab}$ and we apply the York 
decomposition~\cite{J.W.York-1973-1974}: 
\begin{eqnarray}
  {}^{(1)}\!g_{ab}
  &=&
  {}^{(1)}\!h_{\eta\eta}(d\eta)_{a}(d\eta)_{b}
  +
  2 \left(
    D_{p}{}^{(1)}\!h_{(VL)} + {}^{(1)}\!h_{(V)p}
  \right) (d\eta)_{(a}(dx^{p})_{b)}
  \nonumber\\
  &&
  + 
  a^{2}\left\{
    {}^{(1)}\!h_{(L)} \gamma_{pq}
    +
    \left(
      D_{p}D_{q} - \frac{1}{3} \gamma_{pq} \Delta
    \right){}^{(1)}\!h_{(TL)}
  \right.
  \nonumber\\
  && \quad\quad\quad
  \left.
    +
    2 D_{(p}{}^{(1)}\!h_{(TV)q)} + {}^{(1)}\!h_{(TT)pq}
  \right\}(dx^{p})_{a}(dx^{q})_{b}
  ,
  \label{eq:linear-cosmo-Yorks-decomposition}
\end{eqnarray}
where $\Delta:=\gamma^{pq}D_{p}D_{q}$ and $D_{p}$ is the
covariant derivative associated with the metric $\gamma_{pq}$.
Here, ${}^{(1)}\!h_{(V)p}$, ${}^{(1)}\!h_{(TV)p}$, and
${}^{(1)}\!h_{(TT)pq}$ satisfy the properties  
$D^{p}{}^{(1)}\!h_{(V)p}=D^{p}{}^{(1)}\!h_{(TV)p}=0$,
${}^{(1)}\!h_{(TT)pq}={}^{(1)}\!h_{(TT)qp}$, 
${}^{(1)}\!{h_{(TT)}}^{p}_{\;\;p}:=\gamma^{pq}{}^{(1)}\!h_{(TT)pq}=0$, and
$D^{p}{}^{(1)}\!h_{(TT)pq}=0$.


The gauge-transformation rules for the variables 
$\{{}^{(1)}\!h_{\eta\eta}$,
${}^{(1)}\!h_{(VL)}$, ${}^{(1)}\!h_{(V)p}$, ${}^{(1)}\!h_{(L)}$,
${}^{(1)}\!h_{(TL)}$, ${}^{(1)}\!h_{(TV)q}$, ${}^{(1)}\!h_{(TT)pq}\}$
are derived from (\ref{eq:linear-metric-gauge-trans-0}).
Inspecting these gauge-transformation rules, we define the
gauge-variant part ${}^{(1)}\!X_{a}$ in
(\ref{eq:linear-metric-decomp}): 
\begin{eqnarray}
  {}^{(1)}\!X_{a}
  &:=&
  \left(
    {}^{(1)}\!h_{(VL)}
    - 
    \frac{1}{2} a^{2} \partial_{\eta}{}^{(1)}\!h_{(TL)} 
  \right) (d\eta)_{a}
  \nonumber\\
  &&
  +
  a^{2}\left(
    {}^{(1)}\!h_{(TV)p}
    + 
    \frac{1}{2} D_{p}{}^{(1)}\!h_{(TL)} 
  \right) (dx^{p})_{a}
  .
\end{eqnarray}
We can easily check this vector field ${}^{(1)}\!X_{a}$
satisfies (\ref{eq:linear-metric-decomp-gauge-trans}).
Subtracting gauge-variant part ${\pounds}_{{}^{(1)}\!X}g_{ab}$
from ${}^{(1)}\!g_{ab}$, we have the gauge-invariant part
${}^{(1)}\!{\cal H}_{ab}$ in (\ref{eq:linear-metric-decomp}): 
\begin{eqnarray}
  {}^{(1)}\!{\cal H}_{ab}
  &=&
  a^{2}\left\{
    - 2 {}^{(1)}\!\Phi (d\eta)_{a}(d\eta)_{b}
    + 2 {}^{(1)}\!\nu_{p} (d\eta^{})_{(a}(dx^{p})_{b)}
  \right.
  \nonumber\\
  && \quad\quad
  \left.
    + \left(
      - 2 {}^{(1)}\!\Psi\gamma_{pq} + {}^{(1)}\!\chi_{pq}
    \right) (dx^{p})_{a}(dx^{q})_{b}
  \right\}
  ,
\end{eqnarray}
where the properties
$D^{p}{}^{(1)}\!\nu_{p}:=\gamma^{pq}D_{p}{}^{(1)}\!\nu_{q}=0$, 
${}^{(1)}\!\chi_{p}^{\;\;p}:=\gamma^{pq}{}^{(1)}\!\chi_{pq}:=0$,
and $D^{p}{}^{(1)}\!\chi_{qp}=0$ are satisfied.


We have to emphasize that, as shown in
Refs.~\cite{Nakamura:2010yg}, the one to one 
correspondence between the sets of variables
$\{{}^{(1)}\!g_{\eta\eta}$, ${}^{(1)}\!g_{\eta p}$,
${}^{(1)}\!g_{pq}\}$ and $\{{}^{(1)}\!h_{\eta\eta}$,
${}^{(1)}\!h_{(VL)}$, ${}^{(1)}\!h_{(V)p}$, ${}^{(1)}\!h_{(L)}$,
${}^{(1)}\!h_{(TL)}$, ${}^{(1)}\!h_{(TV)q}$,
${}^{(1)}\!h_{(TT)pq}\}$ is guaranteed by the existence of the
Green functions $\Delta^{-1}$, $(\Delta+2K)^{-1}$, and
$(\Delta+3K)^{-1}$. 
In other words, in the decomposition
(\ref{eq:linear-cosmo-Yorks-decomposition}), some perturbative
modes of the metric perturbations which belongs to the kernel of
the operator $\Delta$, $(\Delta+2K)$, and $(\Delta+3K)$ are
excluded from our consideration.
For example, homogeneous modes belong to the kernel of the
operator $\Delta$ and are excluded from our consideration.
If we have to treat these modes, separate treatments are
necessary.
This is the ``zero-mode problem'' in the comsmological
perturbations, which was pointed out in
Refs.~\cite{kouchan-decomp}.


To define gauge-invariant variables for $n$th-order metric
perturbation, we apply the York decomposition
(\ref{eq:linear-cosmo-Yorks-decomposition}) not to the variable
${}^{(n)}\!g_{ab}$ but to the variable ${}^{(n)}\!\hat{H}_{ab}$
defined by (\ref{eq:nth-order-hatHab-def}):
\begin{eqnarray}
  \label{eq:nth-cosmo-Yorks-decomposition}
  {}^{(n)}\!\hat{H}_{ab}
  &=&
  {}^{(n)}\!h_{\eta\eta} (d\eta)_{a}(d\eta)_{b}
  +
  2 \left(
    D_{p}{}^{(n)}\!h_{(VL)} + {}^{(n)}\!h_{(V)p}
  \right) (d\eta)_{(a}(dx^{p})_{b)}
  \nonumber\\
  &&
  + 
  a^{2}\left\{
    {}^{(n)}\!h_{(L)} \gamma_{pq}
    +
    \left(
      D_{p}D_{q} - \frac{1}{3} \gamma_{pq} \Delta
    \right) {}^{(n)}\!h_{(TL)}
  \right.
  \nonumber\\
  && \quad\quad\quad
  \left.
    +
    2 D_{(p}{}^{(n)}\!h_{(TV)q)} + {}^{(n)}\!h_{(TT)pq}
  \right\}(dx^{p})_{a}(dx^{q})_{b}
  .
\end{eqnarray}
Since the gauge-transformation rule
(\ref{eq:nth-hatHab-gauge-trans-final}) for the variable
${}^{(n)}\!\hat{H}_{ab}$ has the same form as the
gauge-transformation rule (\ref{eq:linear-metric-gauge-trans}),
we can define the gauge-variant parts of
${}^{(n)}\!\hat{H}_{ab}$ as
\begin{eqnarray}
  {}^{(n)}\!X_{a}
  &:=&
  \left(
    {}^{(n)}\!h_{(VL)}
    - 
    \frac{1}{2} a^{2} \partial_{\eta}{}^{(n)}\!h_{(TL)} 
  \right) (d\eta)_{a}
  \nonumber\\
  &&
  +
  a^{2}\left(
    {}^{(n)}\!h_{(TV)p}
    + 
    \frac{1}{2} D_{p}{}^{(n)}\!h_{(TL)} 
  \right) (dx^{p})_{a}
\end{eqnarray}
through the same procedure as the linear case and we can also
define the gauge-invariant part ${}^{(n)}\!{\cal H}_{ab}$ by
\begin{eqnarray}
  {}^{(n)}\!{\cal H}_{ab}
  &=&
  a^{2}\left\{
    - 2 {}^{(n)}\!\Phi (d\eta)_{a}(d\eta)_{b}
    + 2 {}^{(n)}\!\nu_{p} (d\eta^{})_{(a}(dx^{p})_{b)}
  \right.
  \nonumber\\
  && \quad\quad
  \left.
    + \left(
      - 2 {}^{(n)}\!\Psi\gamma_{pq} + {}^{(n)}\!\chi_{pq}
    \right) (dx^{p})_{a}(dx^{q})_{b}
  \right\}
  ,
\end{eqnarray}
where the properties
$D^{p}{}^{(n)}\!\nu_{p}:=\gamma^{pq}D_{p}{}^{(n)}\!\nu_{q}=0$, 
${}^{(n)}\!\chi_{p}^{\;\;p}:=\gamma^{pq}{}^{(n)}\!\chi_{pq}:=0$,
and $D^{p}{}^{(n)}\!\chi_{qp}=0$ are satisfied.


As noted in Refs.~\cite{Nakamura:2010yg}, the definitions of
gauge-invariant variables are not unique.
Therefore, we may choose the different choice of gauge-invariant 
variables for each order metric perturbations through the
different choice of ${}^{(n)}\!X_{a}$.
The above choice corresponds to the longitudinal gauge in linear
cosmological perturbations.


\section{Summary and Discussions}
\label{sec:K.Nakamura-2010-6}


In this paper, we discussed the recursive structure in the
construction of gauge-invariant variables for any-order
perturbations.
As gauge-transformation rules for the higher-order
perturbations, we applied the knight diffeomorphism introduced 
by Sonego and Bruni~\cite{S.Sonego-M.Bruni-CMP1998}. 
This diffeomorphism is regarded as general diffeomorphism in the
order-by-order treatment of perturbations.
Based on the gauge-transformation rules for higher-order
perturbations derived by Sonego and
Bruni~\cite{S.Sonego-M.Bruni-CMP1998}, we proposed the procedure
to construct gauge-invariant variables to third order
in~\cite{kouchan-gauge-inv}.
Based on this procedure, in this paper, we consider the explicit
and systematic construction of gauge-invariant variables for
more higher-order perturbations.
As a result, we found that the recursive structure in the
construction of gauge-invariant variables.


Although we do not prove Conjecture
\ref{conjecture:nth-order-gauge-transformation-conjecture}
within this paper, we have confirmed this conjecture to 4th
order.
Therefore, it is reasonable to regard that the algebraic
relation (\ref{nth-hatHab-conjectured-identity-1}) is hold.
Then, the gauge-transformation rule for the variable
${}^{(n)}\!\hat{H}_{ab}$ defined by
equation~(\ref{eq:nth-order-hatHab-def}) is given as
equation~(\ref{eq:nth-hatHab-gauge-trans-final}).
This indicates that we may apply
Conjecture~\ref{conjecture:decomposition-conjecture} to the
variable ${}^{(n)}\!\hat{H}_{ab}$ and we can decompose the metric 
perturbation ${}^{(n)}\!g_{ab}$ of $n$th order into its
gauge-invariant part ${}^{(n)}\!{\cal H}_{ab}$ and the
gauge-variant part ${}^{(n)}\!X^{a}$.
The gauge-transformation rule of the gauge-variant part
${}^{(n)}\!X^{a}$ leads the identity
(\ref{eq:gauge-trans-identity-n-1}) with $i=n$.
The identities (\ref{eq:gauge-trans-identity-n-1})
with $i=1,...,n$ is used when we derived the
gauge-transformation rule for the variable
${}^{(n+1)}\!\hat{H}_{ab}$ which is given by
equation~(\ref{nth-hatHab-gauge-trans-0}) with the replacement
$n\rightarrow n+1$.
Through
Conjecture~\ref{conjecture:nth-order-gauge-transformation-conjecture}
with the replacement $n\rightarrow n+1$, the
gauge-transformation rule for the variable
${}^{(n+1)}\!\hat{H}_{ab}$ is also given in the form
(\ref{eq:nth-hatHab-gauge-trans-final}) with the replacement
$n\rightarrow n+1$. 
Thus, we can recursively construct gauge-invariant variables for
any order perturbations through
Conjectures~\ref{conjecture:decomposition-conjecture} and 
\ref{conjecture:nth-order-gauge-transformation-conjecture}.
In this paper, we have confirmed this recursive structure to
4th order.
This recursive structure is the main point of this paper.


We have to note that
Conjecture~\ref{conjecture:decomposition-conjecture} is highly
nontrivial conjecture, while
Conjecture~\ref{conjecture:nth-order-gauge-transformation-conjecture}
is just an algebraic one.
In~\cite{kouchan-decomp}, we proposed a scenario of a proof of
Conjecture~\ref{conjecture:decomposition-conjecture}.
However, there are missing modes of perturbation in this
scenario which called ``zero modes'' and we also proposed
``zero-mode problem''.
The recursive structure in this paper is entirely based on
Conjecture~\ref{conjecture:decomposition-conjecture}.
Therefore, we have to say that ``zero-mode problem'' is also
essential to the recursive structure in the construction of
gauge-invariant variables for any-order perturbations.


Here, we discuss the correspondence with the recent proposal of
the fully non-linear and exact perturbations by Hwang and
Noh~\cite{J.-c.Hwang-H.Noh-2013}.
Since we can decompose the $n$th-order metric perturbation as
equation~(\ref{eq:nth-order-original-ngab-decomp}), the full
metric (\ref{eq:metric-expansion}), which is pulled back to 
${\cal M}_{0}$ through a gauge ${\cal X}$, is given by 
\begin{eqnarray}
  {\cal X}^{*}_{\lambda}\bar{g}_{ab}
  &=&
  g_{ab}
  +
  \sum_{n=1}^{k}
  \frac{\lambda^{n}}{n!}
  {}^{(n)}\!{\cal H}_{ab}
  \nonumber\\
  &&
  -
  \sum_{n=1}^{k}
  \frac{\lambda^{n}}{n!}
  \sum_{l=1}^{n}
  \frac{n!}{(n-l)!}
  \sum_{\{j_{i}\}\in J_{l}} 
  {\cal C}_{l}(\{j_{i}\})
  {\pounds}_{-{}_{\;{\cal X}}^{(1)}\!X}^{j_{1}}
  \cdots
  {\pounds}_{-{}_{\;{\cal X}}^{(l)}\!X}^{j_{l}}
  {}_{\;\;\;\;\;{\cal X}}^{(n-l)}\!g_{ab}
  \nonumber\\
  &&
  +
  O(\lambda^{k+1})
  .
  \label{eq:full-metric-expansion-original}
\end{eqnarray}
Here, in this equation, the term 
$\displaystyle\sum_{n=1}^{k}\frac{\lambda^{n}}{n!}{}^{(n)}\!{\cal H}_{ab}$
is the gauge-invariant part and the second line is the
gauge-variant part up to $k+1$ order. 
If the right-hand side of
equation~(\ref{eq:full-metric-expansion-original}) converges in
the limit $k\rightarrow\infty$, the limit
$\displaystyle \lim_{k\rightarrow\infty}\sum_{n=1}^{k}\frac{\lambda^{n}}{n!}{}^{(n)}\!{\cal H}_{ab}$  
corresponds to the gauge-invariant variables in the fully
non-linear and exact perturbations proposed by Hwang and
Noh~\cite{J.-c.Hwang-H.Noh-2013}.
The gauge issue of the fully non-linear and exact perturbations
will be justified in this way.


In the case of cosmological perturbations discussed in
Sec.~\ref{sec:K.Nakamura-2010-5}, the components of the
gauge-invariant part 
$\displaystyle\lim_{k\rightarrow\infty}\sum_{n=1}^{k}\frac{\lambda^{n}}{n!}{}^{(n)}\!{\cal H}_{ab}$
for the fully non-linear and exact perturbations are given by
\begin{eqnarray}
  \lim_{k\rightarrow\infty}\sum_{n=1}^{k}\frac{\lambda^{n}}{n!}{}^{(n)}\!{\cal H}_{ab}
  &=&
  a^{2}\left\{
    - 2 {}^{(f)}\!\Phi (d\eta)_{a}(d\eta)_{b}
    + 2 {}^{(f)}\!\nu_{p} (d\eta^{})_{(a}(dx^{p})_{b)}
  \right.
  \nonumber\\
  && \quad\quad
  \left.
    + \left(
      - 2 {}^{(f)}\!\Psi\gamma_{pq} + {}^{(f)}\!\chi_{pq}
    \right) (dx^{p})_{a}(dx^{q})_{b}
  \right\}
  ,
\end{eqnarray}
where
\begin{eqnarray}
  {}^{(f)}\!\Phi
  &:=&
  \lim_{k\rightarrow\infty}\sum_{n=1}^{k}\frac{\lambda^{n}}{n!}
  {}^{(n)}\!\Phi
  , \\
  {}^{(f)}\!\nu_{p} 
  &:=&
  \lim_{k\rightarrow\infty}\sum_{n=1}^{k}\frac{\lambda^{n}}{n!}
  {}^{(n)}\!\nu_{p}
  , \\
  {}^{(f)}\!\Psi
  &:=&
  \lim_{k\rightarrow\infty}\sum_{n=1}^{k}\frac{\lambda^{n}}{n!}
  {}^{(n)}\!\Psi
  , \\
  {}^{(f)}\!\chi_{pq}
  &:=&
  \lim_{k\rightarrow\infty}\sum_{n=1}^{k}\frac{\lambda^{n}}{n!}
  {}^{(n)}\!\chi_{pq}
  .
\end{eqnarray}
However, we have to keep in our mind the fact that we ignored 
``zero modes'' to define the variable ${}^{(n)}\!\Phi$, 
${}^{(n)}\!\nu_{p}$, ${}^{(n)}\!\Psi$, and ${}^{(n)}\!\chi_{pq}$.


Finally, we have to emphasize that the ingredients of this 
paper are also purely kinematical, since the issue of gauge
dependence is purely kinematical. 
Actually, we do not used any information of field equations
such as the Einstein equation. 
Therefore, the ingredients of this paper are applicable to any
theory of gravity with general covariance.


\section*{Acknowledgments}


The author would like to thanks to all members of GW group in
NAOJ for their encouragement.


\appendix
\section{Properties of the set $J_{l}$}
\label{sec:K.Nakamura-2014-appendix}


In \cite{S.Sonego-M.Bruni-CMP1998}, Sonego and Bruni introduced
the set of integer $J_{l}$ associated with the integer $l\geq 1$
defined by
\begin{eqnarray}
  \label{eq:S.Sonego-M.Bruni-1998-examine-1-2}
  J_{l}
  &:=&
  \left\{
    (j_{1},...,j_{n},...)
    \left|
      j_{n}\in\NN,
      \quad \sum_{i=1}^{\infty} ij_{i} = l
    \right.
  \right\}
  \nonumber\\
  \label{eq:1Jl-def}
  &=:&
  {}_{1\!}J_{l}
  ,
\end{eqnarray}
where $\NN$ is the set of natural numbers.
Here, it is convenient to introduce the set $J_{0}$ so that 
\begin{eqnarray}
  \label{eq:introduction-J0}
  J_{0}
  :=
  \left\{
    (j_{1},...,j_{n},...)
    \left|
      j_{n}=0 \quad \forall n\in\NN
    \right.
  \right\}
\end{eqnarray}
Due to this introduction $J_{0}$, we may regard the definition
(\ref{eq:S.Sonego-M.Bruni-1998-examine-1-2}) of $J_{l}$ for
$l\geq 0$.


To classify the elements of $J_{l}$, we first introduce the set  
\begin{eqnarray}
  {}_{1\!}J_{l}^{+}
  :=
  \left\{
    (j_{1}+1,j_{2},...) \left|
      (j_{1},...,j_{l},....) \in {}_{1\!}J_{l}
    \right.
  \right\}.
\end{eqnarray}
We note that 
\begin{eqnarray}
  {}_{1\!}J_{0}^{+} = \{(1,0,0,....)\} = {}_{1\!}J_{1}.
\end{eqnarray}
If we replace $j_{1}\rightarrow j_{1}+1$ in the condition
$\displaystyle\sum_{i=1}^{\infty} ij_{i} = l$ of the definition
(\ref{eq:S.Sonego-M.Bruni-1998-examine-1-2}), we obtain 
\begin{eqnarray}
  j_{1} + \sum_{i=2}^{\infty} ij_{i} = l - 1. 
\end{eqnarray}
Therefore, ${}_{1\!}J_{l-1}^{+}$ is a subset
${}_{1\!}J_{l}$, namely, the elements of ${}_{1\!}J_{l-1}^{+}$
is the elements of ${}_{1\!}J_{l}$ with $j_{1}\geq 1$.
All elements of the set
${}_{1\!}J_{l}\backslash{}_{1\!}J_{l-1}^{+}$ have the property
$j_{1}=0$.


Second, we consider the set
${}_{1\!}J_{l}\backslash{}_{1\!}J_{l-1}^{+}$. 
We define ${}_{2\!}J_{l}^{+}$ by 
\begin{eqnarray}
  {}_{2\!}J_{l}^{+}
  :=
  \left\{
    (j_{1},j_{2}+1,j_{3},...)
    |
    (j_{1},j_{2},j_{3},...)
    \in
    {}_{1\!}J_{l}\backslash{}_{1\!}J_{l-1}^{+}
  \right\}.
\end{eqnarray}
Since all elements in the set
${}_{1\!}J_{l}\backslash{}_{1\!}J_{l-1}^{+}$ have 
the property $j_{1}=0$, all elements in the set
${}_{2\!}J_{l}^{+}$ also have the property $j_{1}=0$.
Furthermore, since the elements in
${}_{1\!}J_{l}\backslash{}_{1\!}J_{l-1}^{+}$ satisfy the
condition $\displaystyle\sum_{i=2}^{\infty} ij_{i} = l$, the
elements of the set ${}_{2\!}J_{l}^{+}$ satisfy the property
$\displaystyle\sum_{i=2}^{\infty} ij_{i} = l+2$. 
This implies that the set ${}_{2\!}J_{l-2}^{+}$ is the subset of
the set ${}_{1\!}J_{l}\backslash{}_{1\!}J_{l-1}^{+}$ with the
property $j_{2}\geq 1$.
We note that all elements of the set
${}_{1\!}J_{l}\backslash\left({}_{1\!}J_{l-1}^{+}\oplus{}_{2\!}J_{l-2}^{+}\right)$ 
have the property $j_{1}=j_{2}=0$.
We also note that ${}_{2\!}J_{1}^{+}$ is an empty set.


Similarly, we consider the set
${}_{1\!}J_{l}\backslash\left({}_{1\!}J_{l-1}^{+}\oplus{}_{2\!}J_{l-2}^{+}\right)$ 
We also define ${}_{3\!}J_{l}^{+}$ by 
\begin{eqnarray}
  {}_{3\!}J_{l}^{+}
  :=
  \left\{
    (j_{1},j_{2},j_{3}+1,j_{4},...)
    |
    (j_{1},j_{2},j_{3},...)
    \in
    {}_{1\!}J_{l}\backslash\left({}_{1\!}J_{l-1}^{+}\oplus{}_{2\!}J_{l-2}^{+}\right)
  \right\}.
\end{eqnarray}
Since all elements in the set
${}_{1\!}J_{l}\backslash\left({}_{1\!}J_{l-1}^{+}\oplus{}_{2\!}J_{l-2}^{+}\right)$
have the property $j_{1}=j_{2}=0$, all elements in the set
${}_{3\!}J_{l}^{+}$ also have the property $j_{1}=j_{2}=0$.
Furthermore, since the elements in
${}_{1\!}J_{l}\backslash\left({}_{1\!}J_{l-1}^{+}\oplus{}_{2\!}J_{l-2}^{+}\right)$
satisfy the condition
$\displaystyle\sum_{i=3}^{\infty}ij_{i}=l$, the  
elements of the set ${}_{3\!}J_{l}^{+}$ satisfy the property
$\displaystyle\sum_{i=2}^{\infty} ij_{i} = l+3$. 
This implies that the set ${}_{3\!}J_{l-3}^{+}$ is the subset of 
the set 
${}_{1\!}J_{l}\backslash\left({}_{1\!}J_{l-1}^{+}\oplus{}_{2\!}J_{l-2}^{+}\right)$
with the property $j_{3}\geq 1$.
We note that all elements of the set
${}_{1\!}J_{l}\backslash\left({}_{1\!}J_{l-1}^{+}\oplus{}_{2\!}J_{l-2}^{+}\oplus{}_{3\!}J_{l-3}^{+}\right)$
have the property $j_{1}=j_{2}=j_{3}=0$ and the sets
${}_{3\!}J_{l}^{+}$ with $l=1,2$ are empty sets.


We can repeat this classification of the elements
in ${}_{1\!}J_{l}$ through the recursive definitions of the sets
\begin{eqnarray}
  {}_{k\!}J_{l}^{+}
  &:=&
  \left\{
    (j_{1},...j_{k-1},j_{k}+1,j_{k+1},...)
    \left|
    (j_{1},...,j_{k},...)
    \in
    {}_{1\!}J_{l}\backslash\left(\bigoplus_{p=1}^{k}{}_{p\!}J_{l-p}^{+}\right)
    \right.
  \right\},
  \nonumber\\
\end{eqnarray}
for $0\geq k \geq l$.
This classification of the elements in ${}_{1\!}J_{l}$
terminates when $k=l$ and we obtain the results 
\begin{eqnarray}
  J_{l} =: {}_{1\!}J_{l} = \bigoplus_{k=1}^{l}{}_{k\!}J_{l-k}^{+}.
\end{eqnarray}
We note that 
\begin{eqnarray}
  {}_{k\!}J_{l-k}^{+} = \emptyset \quad \mbox{for} \quad k > l - k > 0.
\end{eqnarray}
and 
\begin{eqnarray}
  {}_{l\!}J_{0}^{+} = \left\{(0,...,0,j_{l}=1,0,...)\right\}.
\end{eqnarray}


The explicit elements of ${}_{1\!}J_{1}$, ${}_{1\!}J_{2}$,
${}_{1\!}J_{3}$, and ${}_{1\!}J_{4}$ are given by 
\begin{eqnarray}
  {}_{1\!}J_{1} &=& \left\{ (1,0,0,0,0,0....)  \right\}, \\
  {}_{1\!}J_{2} &=& \left\{ (2,0,0,0,0,0....), \right. \nonumber\\
              &&  \;\; \left.  (0,1,0,0,0,0....)  \right\}, \\
  {}_{1\!}J_{3} &=& \left\{ (3,0,0,0,0,0....), \right.\nonumber\\
              &&  \;\; \left.  (1,1,0,0,0,0....), \right.\nonumber\\
              &&  \;\; \left.  (0,0,1,0,0,0....)  \right\}, \\
  {}_{1\!}J_{4} &=& \left\{ (4,0,0,0,0,0....), \right.\nonumber\\
              &&  \;\; \left.  (2,1,0,0,0,0....), \right.\nonumber\\
              &&  \;\; \left.  (1,0,1,0,0,0....), \right.\nonumber\\
              &&  \;\; \left.  (0,2,0,0,0,0....), \right.\nonumber\\
              &&  \;\; \left.  (0,0,0,1,0,0....)  \right\}.
\end{eqnarray}


\section*{References}

\end{document}